\documentclass[fleqn,11pt]{wlscirep}
\hypersetup{final}
\usepackage[utf8]{inputenc}
\usepackage[T1]{fontenc}
\usepackage{bm}
\usepackage{caption}
\usepackage{subcaption}
\usepackage{graphicx}
\usepackage{amsmath}

\title{Inverse designing surface curvatures by deep learning} 

\author[1,2]{Yaqi Guo}
\author[1*]{Saurav Sharma}
\author[1*]{Siddhant Kumar}

\affil[1]{Department of Materials Science and Engineering, Delft University of Technology, Netherlands}
\affil[2]{School of Aerospace Engineering and Applied Mechanics, Tongji University, China}

\affil[*]{S.Sharma-7@tudelft.nl, Sid.Kumar@tudelft.nl}


\newcommand{\boldface}[1]{\boldsymbol{#1}}  % italic (slanted)

\newcommand{\bfe}{\boldface{e}}

\newcommand{\bfv}{\boldface{v}}

\newcommand{\bfx}{\boldface{x}}

\newcommand{\bfM}{\boldface{M}}

\newcommand{\bfkappa}{\boldsymbol{\kappa}}

\newcommand{\bfchi}{\boldsymbol{\chi}}

\newcommand{\bfTheta}{\boldsymbol{\Theta}}

\newcommand{\calD}{\mathcal{D}}

\newcommand{\calF}{\mathcal{F}}
\newcommand{\calG}{\mathcal{G}}

\newcommand{\calS}{\mathcal{S}}

\newcommand{\calU}{\mathcal{U}}

\newcommand{\Rset}{\mathbb{R}}

\newlength{\boxwidth}
\def\dd{\;\!\mathrm{d}}

\def\btheorem{\begin{theorem}}
\def\etheorem{\end{theorem}}
\def\blemma{\begin{lemma}}
\def\elemma{\end{lemma}}
\def\bproposition{\begin{proposition}}
\def\eproposition{\end{proposition}}
\def\bcorollary{\begin{corollary}}
\def\ecorollary{\end{corollary}}
\def\bdefinition{\begin{definition}}
\def\edefinition{\end{definition}}
\def\bexample{\begin{example}}
\def\eexample{\end{example}}
\def\bremark{\begin{remark}}
\def\eremark{\end{remark}}

\newcommand{\be}{\begin{equation}}
\newcommand{\ee}{\end{equation}}
\newcommand{\beq}{\begin{eqnarray}}
\newcommand{\eeq}{\end{eqnarray}}
\newcommand{\bem}{\begin{multline}}
\newcommand{\eem}{\end{multline}}
\newcommand{\ba}{\begin{align}}
\newcommand{\ea}{\end{align}}

\renewcommand{\figurename}{Figure}
\newcommand{\figurenames}{Figures}
\renewcommand{\tablename}{Table}

\begin{abstract} 
Smooth and curved microstructural topologies found in nature -- from soap films to trabecular bone -- have inspired several mimetic design spaces for architected metamaterials and bio-scaffolds. However, the design approaches so far have been ad hoc, raising the challenge: how to systematically and efficiently inverse design such artificial microstructures with targeted topological features? Here, we explore surface curvature as a design modality and present a deep learning framework to produce topologies with as-desired curvature profiles. The inverse design framework can generalize to diverse topological features such as tubular, membranous, and particulate features. Moreover, we demonstrate successful generalization beyond both the design and data space by inverse designing topologies that mimic the curvature profile of trabecular bone, spinodoid topologies, and periodic nodal surfaces for application in bio-scaffolds and implants. Lastly, curvature and mechanics are bridged by showing how topological curvature can be designed to promote mechanically beneficial stretching-dominated deformation over bending-dominated deformation.
\end{abstract}

\begin{document}

\flushbottom
\maketitle

\thispagestyle{empty}

\section{Introduction}

\begin{figure}
	\centering
	\includegraphics[width=\linewidth]{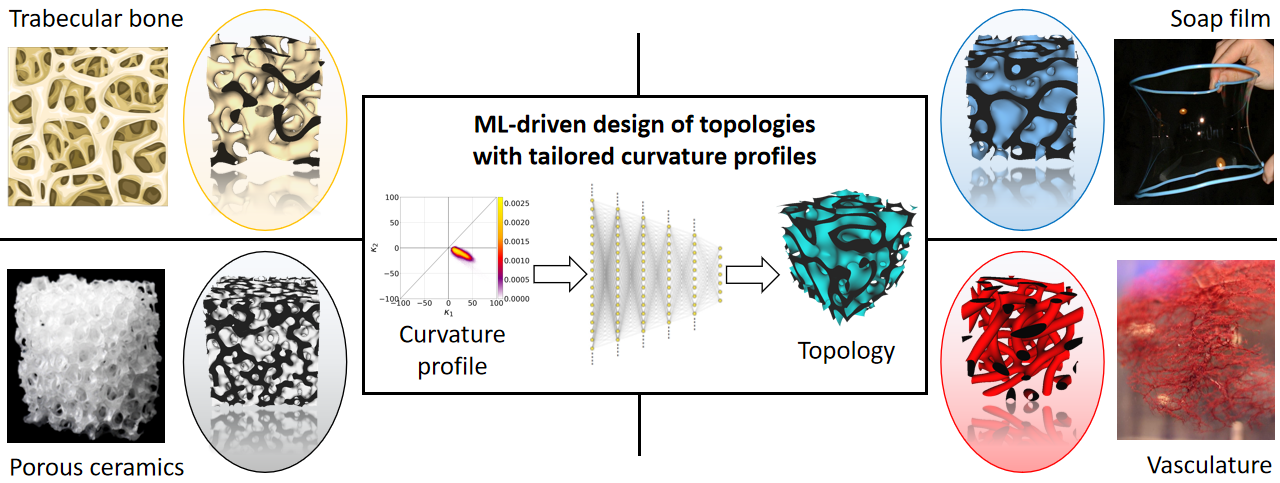}
	\caption{Natural materials and structures such as trabecular bone, soap films, porous ceramics, and vascular systems tend to have smooth curved topologies. Borrowing inspiration from those topologies, the ML-based inverse design framework provides a unified solution for on-demand rational design or mimicry of various smooth topologies with targeted curvature profiles. Since the curvature profile of a topology is directly related to its mechanical properties, this approach provides an efficient pathway to designing mechanical metamaterials with superior properties for applications in light-weight engineering materials, bio-mimetic structures, bio-implants, and more. {Images adapted: 
			\textit{trabecular bone} by \textit{Laboratoires Servier}, CC BY-SA 3.0, via Wikimedia Commons;\cite{bone}
			\textit{soap film}  by \textit{Blinking Spirit}, CC0, via Wikimedia Commons;\cite{soapfilm}
			\textit{porous ceramics}  by \textit{Onnovisser1979}, CC BY-SA 3.0, via Wikimedia Commons;\cite{ceramic}
			\textit{vasculature} by \textit{I'm in the garden}, CC BY-SA 3.0, via Wikimedia Commons.\cite{vessels} }}
	\label{fig:abstract}
\end{figure}

Nature is replete of porous structures with unique curvature topologies -- from the simplest example of soap films with constant mean curvatures (Plateau's law \cite{taylor1976,almgren1976}) to the morphogenesis-driven Turing patterns \cite{Turing1952,Kondo2017} in animal skin pigmentation; from biological systems such as trabecular bone \cite{callens2021,Ralph2009,Puhka2012} and vascular networks \cite{ramasamy2017,sivaraj2016} to non-biological systems like porous ceramics, nanoporous gold \cite{Erlebacher2001}, and block copolymers \cite{portela2020extreme} (see \figurename~\ref{fig:abstract}). Driven by complex relaxation dynamics and non-equilibrium phenomena such as self-assembly,\cite{hill2019,jung2019,ren2018} pattern formation,\cite{paul2021,jiang2021,anzaki2021} and phase ordering kinetics,\cite{diaz2022,ankudinov2022} both understanding and tuning such physics behind the natural emergence of complex curvature topologies opens up new avenues for advances in materials engineering.

For instance, mimicking the microstructural curvature topologies of nanoporous gold and phase-separated block copolymers, spinodoid metamaterials  \cite{Soyarslan2018,Meng2019,portela2020extreme,kumar2020} have  sparked significant interest for their potential applications in bio-implants,\cite{Meng2021} lightweight structures,\cite{zheng2021data} energy absorption,\cite{GuellIzard2019} mass transport,\cite{Roding2022} and more. The microstructure of these materials emerges from a spinodal decomposition process (i.e., rapid  separation of immiscible phases) via either self-assembly \cite{portela2020extreme} or combination of in-silico design and 3D-printing.\cite{Meng2019} In spinodal decomposition, the phase separation is governed by minimization of the bulk free energy while regulated by the interfacial energy between the phases; the latter being intricately related to the interfacial curvatures. The resulting curvature topology promotes stretching-dominated deformation, which is stronger than bending-dominated deformation; therefore, exhibits excellent mechanical resilience.\cite{portela2020extreme,kumar2020} However, how to (in silico) tune the physics of spinodal decomposition to obtain tailored curvature topologies for favorable mechanics remains an open question.

A similar curvature design challenge is gaining attention in the field of bio-scaffolds and implants. Substrate curvatures play a significant role in inducing spatiotemporal growth, differentiation, and migration of biological cells and tissues.\cite{lou2022,Werner2016,zadpoor2015,zhang2022,Sebastien2020,Callens2023} For example, metamaterials based on triply periodic minimal surfaces (TPMS) gained popularity in the design of bone implants as they were believed to mimic the zero mean curvature of trabecular bone and in turn, enhance the long-term compatibility. However, a recent study \cite{callens2021} has challenged this assumption, revealing that trabecular bone displays a complex curvature profile instead, with significant variations across patients and anatomical sites. This highlights the need to develop porous structures with tunable curvature topologies, not only for bone implants but also for cell scaffolds that can adapt to diverse patient- and (anatomical) site-specific contexts.

Despite a wide variety of explorations ranging from TPMS to spinodoid metamaterials, a more general and unified topology description of smooth porous microstructures remains to be investigated. In this direction, Song \cite{Song2021} recently proposed a unifying phase-field framework to generate tubular and membranous topologies with diverse curvature profiles. Similar to the canonical spinodal decomposition model\cite{Cahn1958}, topologies within this framework are obtained as optimizers (subject to constant volume restriction) of energy functionals parameterized based on principal curvatures of the phase-field interface. While this approach opens up a design space to control curvature topology, key challenges persist. \textit{(i)}~Exploring the entire design-property space is nearly impossible because each topology requires a computationally expensive phase-field simulation. The characterization of the microstructural curvature profile by a high-dimensional probability distribution of principal curvatures further exacerbates the situation. \textit{(ii)}~Although the curvature profile of a given topology is unique and simple to obtain, the inverse design, i.e.,  identifying a microstructural topology with a targeted curvature profile remains an open challenge. This is due to the ill-posedness of the inverse problem, i.e., multiple topologies with different design parameters may display the identical or similar curvature profiles, making it difficult to identify the optimal design. In addition, traditional design approaches such as topology optimization \cite{Sigmund2013, zhai2024topology, gao2020comprehensive} and intuition-based approaches are not only computationally expensive, but they may also not be applicable to design for an unconventional target property, such as the microstructural curvature profile. To enhance the discovery of novel topologies, traditional empirical methods typically rely on collecting data from experiments or drawing inspiration from nature to mimic biological materials.\cite{gibson2010cellular, espinosa2011tablet, barthelat2007experimental, barthelat2011toughness} While intuition-based design approaches can be valuable, they often encounter limitations stemming from a restricted exploration of design possibilities. By relying solely on intuition, designers may overlook alternative perspectives and innovative solutions, leading to a narrower design space. For instance, while TPMS with varying Gaussian curvatures can be designed and fabricated\cite{yang2022gaussian}, the parameterization lacks generalizability to a broad and diverse range of smooth surfaces with tailored curvature profiles.

To overcome this limitation, a machine learning (ML) framework is introduced for generating smooth porous microstructures with inverse-designed curvature profiles (see \figurename~\ref{fig:abstract}). 
By analyzing vast amounts of experimental and computational data, ML algorithms can bypass inefficient trial-and-error methods and efficiently unravel hidden high-dimensional structure-property relations, leading to the design of novel materials with tailored properties. Employing a diverse array of ML techniques -- from classical Bayesian optimization \cite{garnett_bayesoptbook_2023} to modern approaches such as variational autoencoders,\cite{PinheiroCinelli2021} generative adversarial networks,\cite{Goodfellow2014} and diffusion models\cite{yang2023diffusion} -- recent applications span a wide range of domains, including mechanical 
metamaterials,\cite{kumar2020,Bastek2021,Wang2020,Bessa2019,vlassis2023denoising,Mao2020} composites,\cite{Gu2018}, photonics,\cite{Malkiel2018} biomaterials,\cite{Suwardi2021} and more. In the context of metamaterials and porous materials,  while mechanical properties such as stiffness, Poisson's ratio, buckling modes, diffusivity, wave propagation, etc. have been inverse-designed, microstructural curvature as a design modality has not been explored. 

To this end, we enable instant identification of microstructural topologies with precise targeted curvature profiles and unlock new possibilities in design of metamaterials and porous microstructures. This is achieved by strategically using a dual deep neural network (NN) setup \cite{kumar2020,Bastek2021,VANTSANT2023} in tandem with a phase-field framework driven by curvature-based energetics \cite{Song2021} that bypasses both the aforementioned key challenges -- namely, the computational bottleneck of phase-field methods and the ill-posedness in inverse design.  

With the ability to generate tailored microstructural curvature profiles, we further investigate the interlink between curvature and mechanics. The membrane-flexural coupling appearing in the mechanics of shell/curved structures is strongly linked with their principal curvatures  \cite{reddy2006theory}, i.e., the redistribution of applied loads in the form of bending and stretching/compressive loads depends on the two principal curvatures of a curved surface. For a selection of micro-architectures with diverse curvature profiles, this load redistribution is studied by decomposing the stored strain energy into membrane and bending strain energies. Since a dominance of membrane deformation over bending deformation is known to promote higher mechanical resilience, the study of ratios of membrane to bending strain energies for diverse curvature profiles can guide the design of superior mechanical metamaterials through the presented inverse design approach.

In the following, we first introduce the design space and discuss the phase-field approach for topology generation. Subsequently, the ML framework for inverse design is introduced. Next, we demonstrate successful inverse design generalization for target curvatures beyond those in the training domain and show application to the design of topologies that mimic trabecular bone, spinodoid surfaces, and nodal surfaces. Lastly, mechanical aspects of the curvature based microstructures are presented by investigating a set of topologies with diverse curvature profiles for their distribution of membrane and bending strain energies.

\section{Results and Discussion}

\subsection{Design space from curvature-based energy functional}

\begin{figure}
	\centering
	\includegraphics[width=\linewidth]{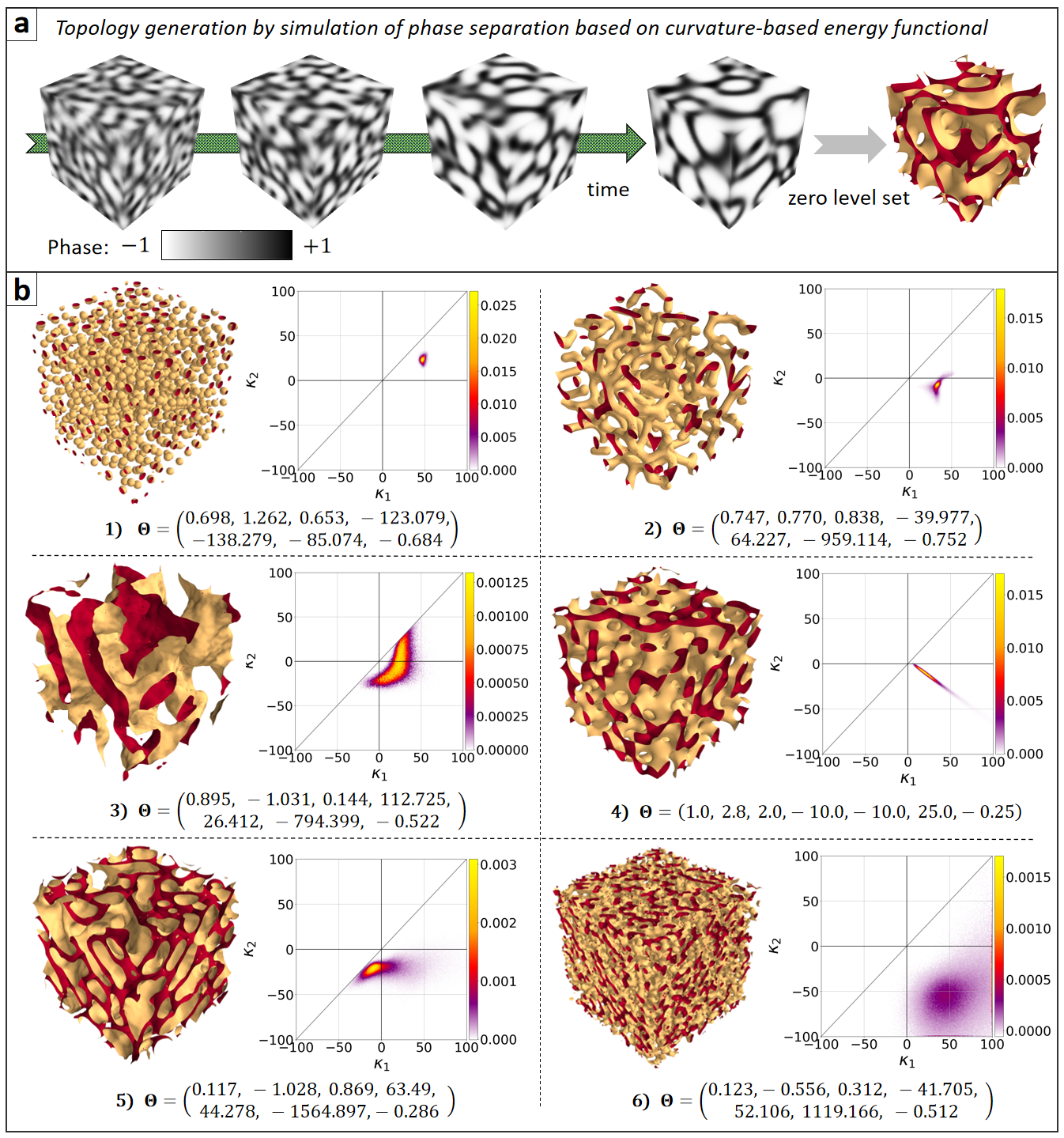}
	\caption{\textbf{(a)} Schematic of topology generation using computationally expensive phase-field simulation based on curvature-driven energy functional. \textbf{(b)} Representative selection of diverse topologies and their corresponding curvature profiles for different design parameters $\bfTheta$. Typical microstructural features include spheres, tubules, or membranes. The curvature profile is visualized as the density scatter of the (surface) element-wise principal curvatures  ($\kappa_1$ and $\kappa_2$); the density color of each scatter point is proportional to the cumulative surface area of the elements with similar principal curvatures in the mesh. Two elements are said to have the similar principal curvatures if they lie in the same cell of a finely-gridded $\kappa_1$-$\kappa_2$ plane.}
	\label{fig:topology-generation}
\end{figure}

We adopt the curvature-based phase-field framework of Song \cite{Song2021} which makes it possible to unify topologies with spherical, tubular, and membranous features as well as their combinations with seamless transition.  Briefly, for a given surface {$\calS$}, let us consider a energy functional $F$ based on the surface curvatures as
\be
\label{eq:F}
F[\calS] = \int_\calS f[S] \dd S \qquad \text{with} \quad f[S] = a_{20}\kappa_1^2+a_{11}\kappa_1\kappa_2+a_{02}\kappa_2^2+a_{10}\kappa_1+a_{01}\kappa_2+a_{00},
\ee
where the integrand $f[S]$ (i.e., surface energy density) is a second-degree polynomial of the principal curvatures ${\kappa_1,\kappa_2}$ (by convention, $\kappa_1\geq \kappa_2$) at the point $S$ on the surface $\calS$ and parameterized by the coefficients $\{a_{20},a_{11},a_{02},a_{10},a_{01},a_{00}\}$. Similar to energy functional of the canonical Cahn-Hilliard equation \cite{Cahn1958,Vidyasagar2018,kumar2020}, minimizing this energy functional represents a compromise between minimizing the bending energy vs.~the surface area. The energy functional is minimized under constant volume constraint and periodic boundary conditions to yield diverse topologies.  The energy minimization can be approximated by a phase-field model using a mass-preserving ${H^{-1}}$ gradient flow of a two-phase system  as
\be\label{eq:u}
\dot{u} = \Delta\frac{\partial F_\epsilon}{\partial u}(u) \qquad \text{with} \quad m_0 = \frac{1}{|\Omega|}\int_\Omega u \dd V,
\ee
where $u:\Omega \rightarrow [-1,1]$ denotes the phase field, i.e., a concentration field of one phase in the domain $\Omega$. Without loss of generality, we assume a non-dimensionalized domain $\Omega = [0,100]^3$ (consequently, all subsequent length dimensions are normalized relative to the domain size). We assume the phase field contains diffused interfaces with a hyperbolic tangent profile and thickness lengthscale of $\epsilon\ll 1$. ${F_\epsilon}$ is the corresponding phase-field approximation of $F$. $m_0$ denotes the constant volume constraint. $u$ is initialized with  random noise. After solving the gradient flow problem, the final surface $\calS$ is extracted by applying the zero level set on the resulting phase field. The zero level set is not obtained exactly but rather approximated as a triangular surface mesh  $\hat \calS$  by using the marching cubes algorithm \cite{Lorensen1987}. \figurename~\ref{fig:topology-generation}a shows a schematic of the topology generation process. 

Within the phase-field approximation, we speak of diffused curvatures (or curvatures of diffused interfaces). Since the phase field ${u}$ has a hyperbolic tangent profile, the diffused principal curvatures correspond precisely to the respective principal curvatures \cite{Goldman2005} at level sets of ${u}$. Detailed mathematical proof of phase-field approximation as well as implementation details can be found in the ref. \cite{Song2021}.

Let ${\bfTheta=(a_{20},a_{11},a_{02},a_{10},a_{01},a_{00},m_0)}^\intercal$ be a vector of the design parameters that uniquely characterize the gradient flow and hence, the topology generation. Note that the topologies are still stochastic due to the random initial conditions. \figurename~\ref{fig:topology-generation}b shows a representative selection of  diverse topologies and their corresponding curvature profiles obtained for different values of $\bfTheta$. 

For the curvature profile to be interpretable to the NN-based design algorithm, an encoding of the curvature profile is created for a given meshed surface~$\hat \calS$. The probability of $\hat \calS$ containing the principal curvatures $(\kappa_1,\kappa_2)$ is defined as 
\be
p(\kappa_1,\kappa_2 | \hat \calS) =  \frac{\sum_{i=1}^e \delta\big((\kappa_1,\kappa_2),( \hat \kappa_{1,i}, \hat\kappa_{2,i})\big) \hat a_i}{\sum_{i=1}^e  \hat a_i},
\ee
where $\delta$ denotes the Kronecker delta, $e$ is the total number of elements, and $\hat a_i$ and $(\hat \kappa_{1,i}, \hat\kappa_{2,i})$ denote the area and principal curvatures of the $i^\text{th}$ element in the mesh, respectively. Next, this probability distribution in $\kappa_1$-$\kappa_2$ is discretized on a uniform grid of $200\times200$ bins (i.e., a two-dimensional histogram) yielding a $[0,1]^{200\times 200}$ matrix encoding of the curvature profile. Note that this discrete probability matrix is lower-triangular due to the convention of $\kappa_1\geq\kappa_2$. To reduce the computational cost, the lower-triangular part is serialized into a $k=20,100$-dimensional encoding $\bfchi\in[0,1]^{k}$, which is used in the downstream ML tasks. 

\subsection{Geometric interpretation of the design space}

\begin{figure}
	\centering
	\includegraphics[width=\textwidth]{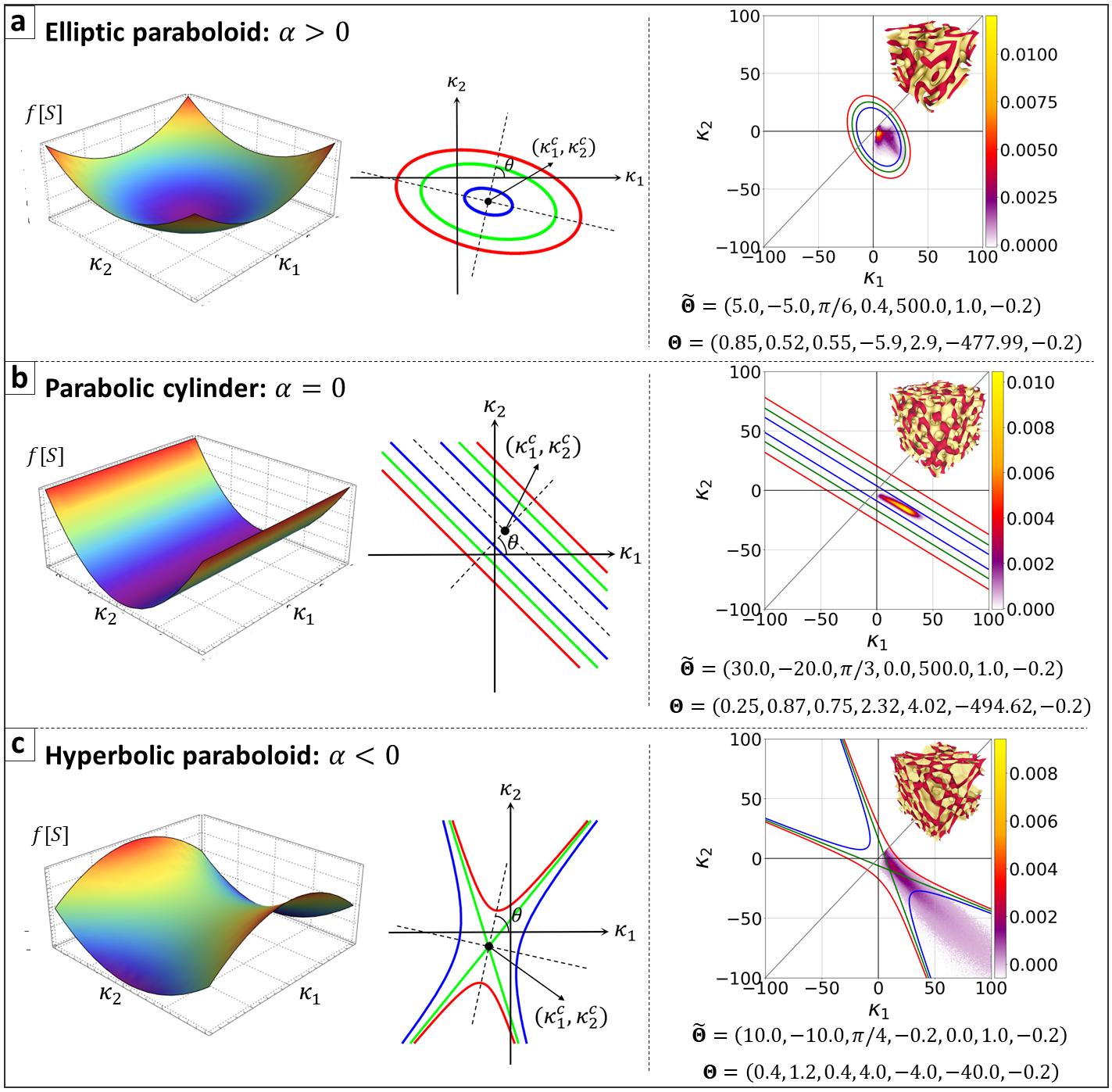}
	\caption{Geometric interpretation of the design space. For the cases when \textbf{(a)} $\alpha>0$, \textbf{(b)} $\alpha=0$, and \textbf{(c)} $\alpha<0$, the surface energy density $f[S]$ can be visualized as different kinds of quadratic surfaces (left column) and the corresponding contours (middle column) of $\kappa_1$-$\kappa_2$. For each case, a representative example (right column)  shows the qualitative correspondence between relevant contours of $f[S]$ and the resulting curvature profile.} 
	\label{fig:geometry}
\end{figure}

The parameters $\{a_{20},a_{11},a_{02},a_{10},a_{01},a_{00}\}$ in $\bfTheta$ admit a characteristic geometric interpretation that directly correlates to the curvature profile of the resulting microstructure. The surface energy density in \eqref{eq:F} can be reformulated as a quadratic surface in the $\kappa_1$-$\kappa_2$-$f$ space such that
\be
f[S] = g \left( 
\tilde\bfkappa^\intercal \bfM \tilde\bfkappa - c\right)\qquad \text{with} 
\quad \tilde\bfkappa = 
\underbrace{\begin{pmatrix}
		\cos\theta & \sin\theta\\
		-\sin\theta & \cos\theta\\
\end{pmatrix}}_{\text{rotation}}
\underbrace{\begin{pmatrix}
		\kappa_1 - \kappa_1^c \\
		\kappa_2 - \kappa_2^c
\end{pmatrix}}_{\text{translation}}
\quad \text{and} \quad
\bfM = \underbrace{\begin{pmatrix}
		1 & 0 \\
		0 & \alpha
\end{pmatrix}}_{\text{aspect ratio}}.
\ee
Here, $(\kappa_1^c,\kappa_2^c)^\intercal\in\Rset^2$, $\theta\in[-\pi/2,\pi/2)$, and $\alpha\in \Rset$ represent the translation, (counter-clockwise) rotation, and aspect ratio in the $\kappa_1$-$\kappa_2$ plane, respectively. The parameters $c\in\Rset$ and $g\in\Rset^{+}$ denote a vertical translation or bias along $f$ and  scaling factor, respectively; the latter does not affect the interrelations between the individual curvature terms of the overall energy density. Supporting Information Section S1 provides conversion formulas between the design parameters $\bfTheta$ and the equivalent geometrically interpretable ones, i.e., $\tilde\bfTheta=(\kappa_1^c,\kappa_2^c,\theta,\alpha,c,g,m_0)^\intercal$. In the cases when $\alpha$ is positive, zero, and negative, the quadratic surface is respectively an elliptic paraboloid, parabolic cylinder, and hyperbolic paraboloid, and consequently, produce elliptic, linear, and hyperbolic contours on the $\kappa_1$-$\kappa_2$ plane. \figurename~\ref{fig:geometry} illustrates how the combination of rotation, translation, and aspect ratios yields a rich design space of surface energy densities.

The geometric interpretation of the surface energy density is reflected in the curvature profiles as the density plots are aligned with the contours of the quadratic surface of the surface energy density (see \figurename~\ref{fig:geometry} for representative examples). This is expected as the hot spots in the density plots qualitatively correspond to the regions of low surface energy density with respect to the principal curvatures. 

\subsection{Data-driven inverse design}

\begin{figure}[t]
	\centering
	\includegraphics[width=\textwidth]{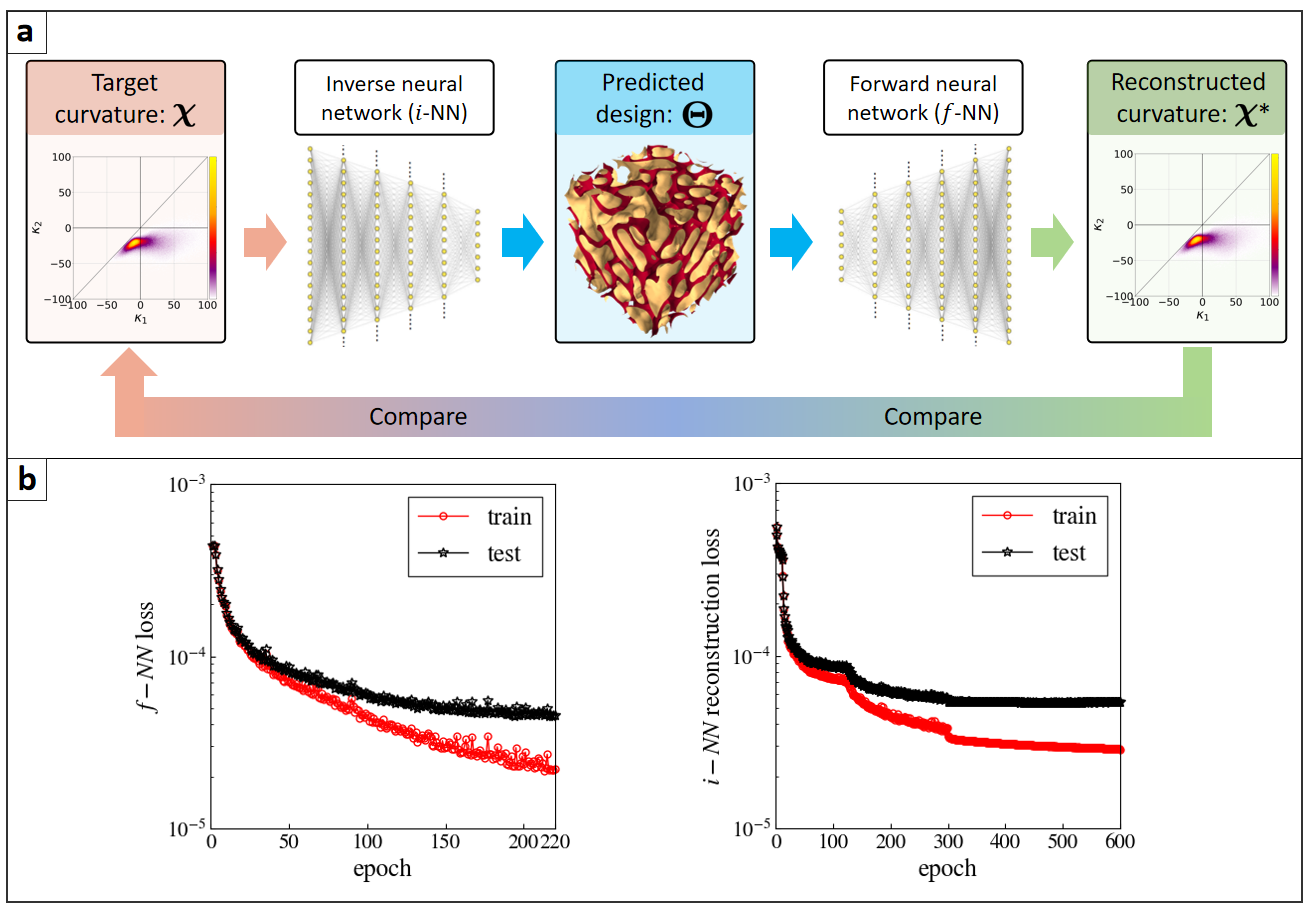}
	\caption{\textbf{(a)} Schematic of the ML-driven inverse design. For a target curvature encoding $\bfchi$, the inverse model ${i}$-NN outputs the design parameters $\bfTheta$. The pre-trained forward model ${f}$-NN then bypasses the evolution of the phase field based on the energetics defined by $\bfTheta$ and directly reconstructs the curvature encoding $\bfchi^*$.  The difference between the target and reconstructed curvature encodings is then used to train the ${i}$-NN. \textbf{(b)} Epoch-wise loss values (evaluated on the training and test datasets separately)  during iterative training of the \textit{f}-NN and \textit{i}-NN.}
	\label{fig:nn}
\end{figure}

In light of the above relation between the design parameters and the curvature profile, we aim to invert this design-property space, i.e., identify the optimal $\bfTheta$ to achieve a target curvature profile. However, as discussed earlier, there are two key challenges: \textit{(i)} the phase-field approach for topology generation is computationally expensive and therefore, a trial-and-error approach to inverse design is intractable; \textit{(ii)}  the inverse design problem is ill-posed as multiple designs can have similar curvature profiles. To this end, we introduce a dual deep-NN setup (see Figure~\ref{fig:nn}a for a schematic). A forward NN first surrogates the phase-field framework and maps the design parameters to the resulting microstructural curvature profile. An inverse NN then inverts the above design-to-curvature map and outputs the appropriate design parameters (and corresponding phase-field energetics) required for a targeted curvature profile.

We start by creating a representative \textit{training} dataset $\calD=\{\{\bfTheta^{(i)},\bfchi^{(i)}\},i=1,\dots,n\}$ consisting of $n=18,000$ pairs of design parameters $\bfTheta$ and the corresponding curvature profiles encoded as $\bfchi$. Since the probability encodings can be sparse and skewed in distribution, independent nonlinear scaling is performed on the individual components of $\bfchi$ to enhance the sensitivity of the downstream ML approach. Supporting Information Section S2 provides further details on the aforementioned data sampling and scaling strategy.

Let $\calF_\omega:\Rset^7\rightarrow \Rset^k$ denote a forward NN (\textit{f}-NN) with a multi-layer perceptron architecture parameterized by the set of trainable weights and biases $\omega$. \textit{f}-NN bypasses the phase-field evolution and directly maps the design parameters $\bfTheta$ to the curvature encoding $\bfchi=\calF_\omega[\bfTheta]$. Since each design has a unique curvature profile (in the sense of averaging across the stochastic effects in topology generation), the forward problem is well-posed. Therefore, we train the \textit{f}-NN by minimizing the (mean squared error) loss between the true and predicted curvature encodings with respect to the \textit{f}-NN-parameters $\omega$, i.e.,
\be\label{eq:fnn}
\calF_\omega \leftarrow \min_\omega \frac{1}{n} \sum_{i=1}^n \|\calF_\omega[\bfTheta^{(i)}] - \bfchi^{(i)} \|^2.
\ee

We now tackle the inverse challenge. Let $\calG_\tau:\Rset^k\rightarrow \Rset^7$ denote an inverse NN (\textit{i}-NN) with a multi-layer perceptron architecture parameterized by the set of trainable weights and biases $\tau$. The \textit{i}-NN maps an input target curvature encoding $\bfchi$ to the design parameters $\bfTheta=\calG_\tau[\bfchi]$. However, unlike the \textit{f}-NN, a training strategy analogous to  \eqref{eq:fnn} fails due to the ill-posedness of the inverse problem. E.g., a high value of the naive loss $\|\calG_\tau[\bfchi] - \bfTheta \|^2$ may indicate that the design parameters predicted by the \textit{i}-NN and the ones from the dataset are different but does not hint anything about the possibility that these dissimilar designs may have similar curvature profiles. To counter this issue, we train the \textit{i}-NN against the pre-trained \textit{f}-NN as
\be\label{eq:inn}
\calG_\tau \leftarrow \min_\tau \frac{1}{n} \sum_{i=1}^n \|\calF_\omega[\calG_\tau[\bfchi^{(i)}]] - \bfchi^{(i)} \|^2.
\ee
For the design predicted by the \textit{i}-NN, i.e., $\calG_\tau[\bfchi^{(i)}]$, the \textit{f}-NN reconstructs the curvature encoding $\bfchi^* = \calF_\omega[\calG_\tau[\bfchi^{(i)}]]$; the loss of $\bfchi^*$ relative to the target $\bfchi$ from the dataset is minimized with respect to the \textit{i}-NN parameters $\tau$. The advantage of this loss function and training strategy over the naive approach is demonstrated below and in Supporting Information Section 3. Note that the ML model is independent of the resolution and scale of the structure, as it takes the probability distribution of the point-wise surface curvatures (referred to as curvature encodings) as its input and outputs the design parameters for the phase-field model.

Both NNs are trained using gradient-based optimization. During \textit{i}-NN training, the \textit{f}-NN not only surrogates the computationally expensive phase-field evolution for curvature reconstruction, but also provides a differentiable map between $\bfTheta$ and $\bfchi$. This differentiability is critical to computing the gradient $\partial \calF_\omega/\partial \bfTheta$ which is in turn required for computing the sensitivity of the loss function in \eqref{eq:inn} with respect to the  \textit{i}-NN parameters $\tau$ during training.
Supporting Information Section S3 provides additional implementation details as well as computational costs associated with both the NNs.

\begin{figure}
	\includegraphics[width=\textwidth]{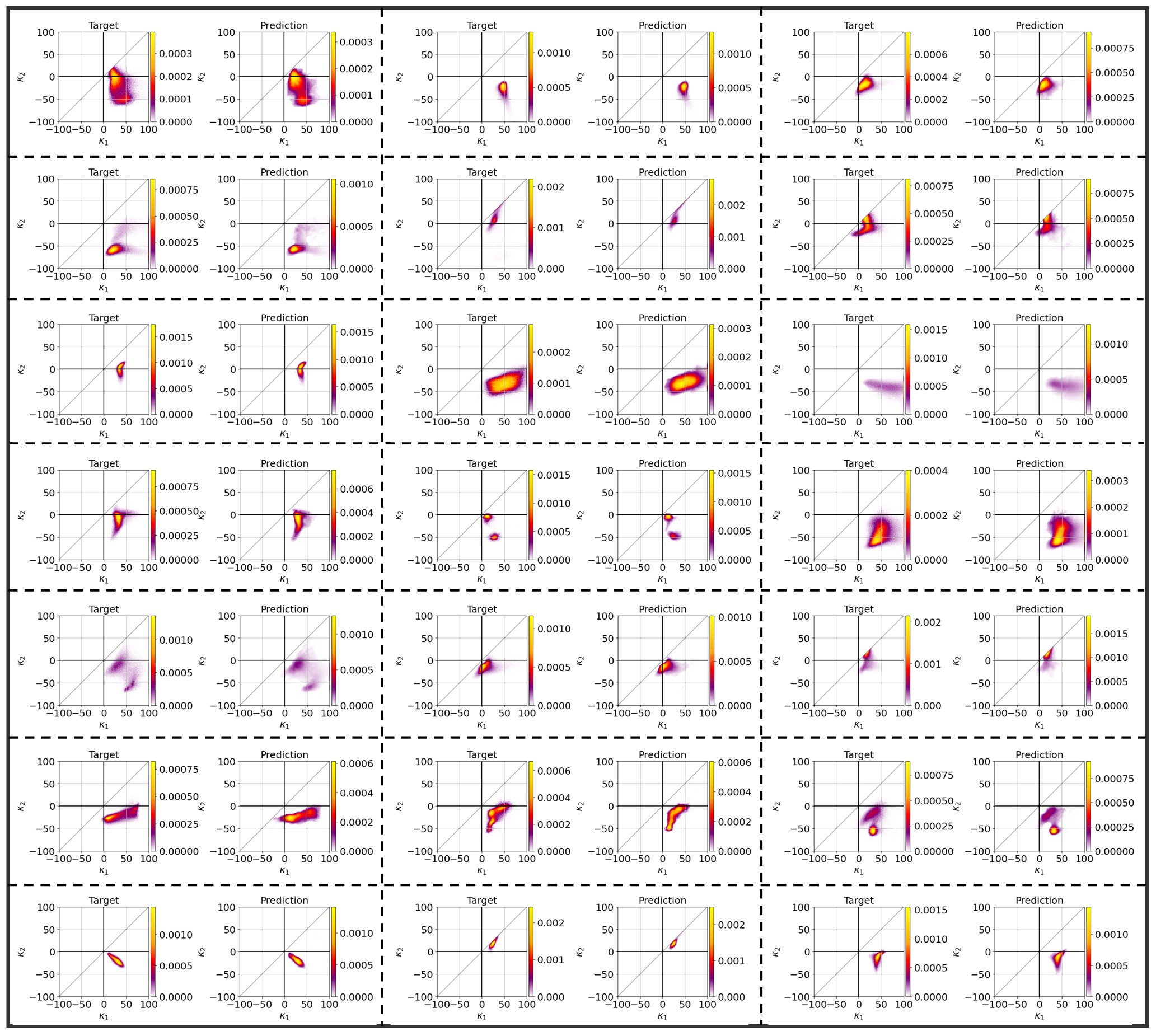}
	\caption{A gallery of representative examples of targeted curvature profiles from the previously unseen test dataset vs. curvature profiles of the topologies predicted by the \textit{i}-NN.}
	\label{fig:gallery}
\end{figure}

\begin{figure}
	\includegraphics[width=\textwidth]{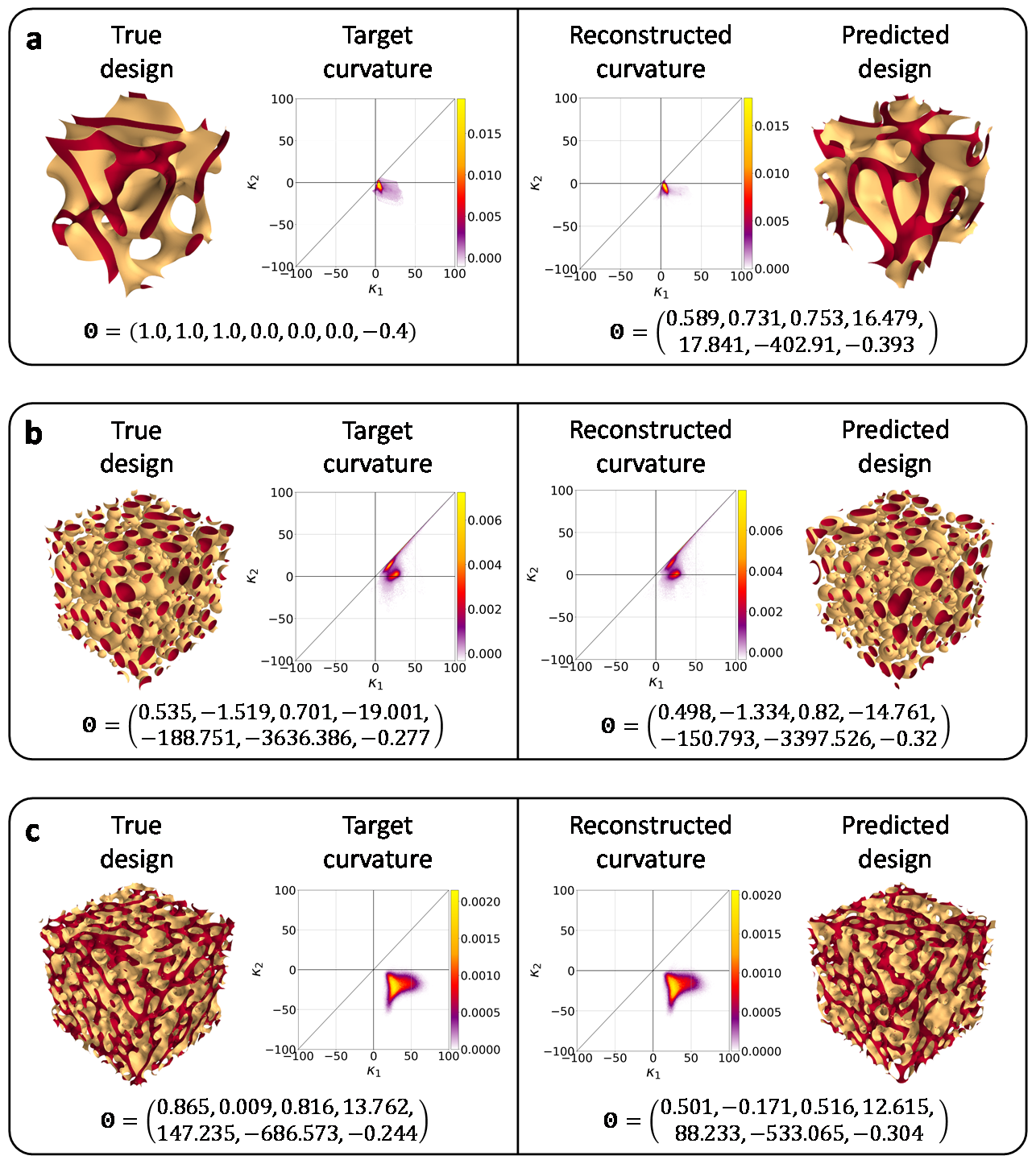}
	\caption{Comparisons between the design parameters $\bfTheta$ (and the resulting topologies) in training dataset and the ones predicted by the \textit{i}-NN.}
	\label{fig:illposed}
\end{figure}

To evaluate the performance of the inverse design framework, we create an additional test dataset of $2,000$ design parameters and curvature encoding pairs, previously not seen by either of the NNs during training. \figurename~\ref{fig:nn}b shows that both \textit{f}-NN and \textit{i}-NN achieve high accuracy (indicated by the low loss values) across both training and test datasets without the presence of any under-/overfitting.
A gallery of 21 representative examples in \figurename~\ref{fig:gallery} shows exceptional agreement in the target curvature profiles  versus the curvature profiles of the designs predicted by the \textit{i}-NN, thereby  providing qualitative evidence of the accuracy of the inverse design approach. \figurename~\ref{fig:illposed}  shows  three additional examples along with a comparison of the design parameters $\bfTheta$ (and the resulting topology) in the dataset and the ones predicted by the \textit{i}-NN. While the target and reconstructed curvature profiles are in agreement with each other, the design parameters are significantly different. It is further pronounced in the  \textit{expectedly} poor correlation between the true design parameters of the test dataset and the predicted design parameters from the \textit{i}-NN; see Supporting Information Section S3 Figure S4. This verifies the ill-posedness of the inverse problem (i.e., multiple designs parameters can lead to similar curvature profiles) and the advantage of the \textit{i}-NN training strategy presented in \eqref{eq:inn}.

\subsection{Generalization beyond the training space}

\begin{figure}[t]
	\includegraphics[width=\textwidth]{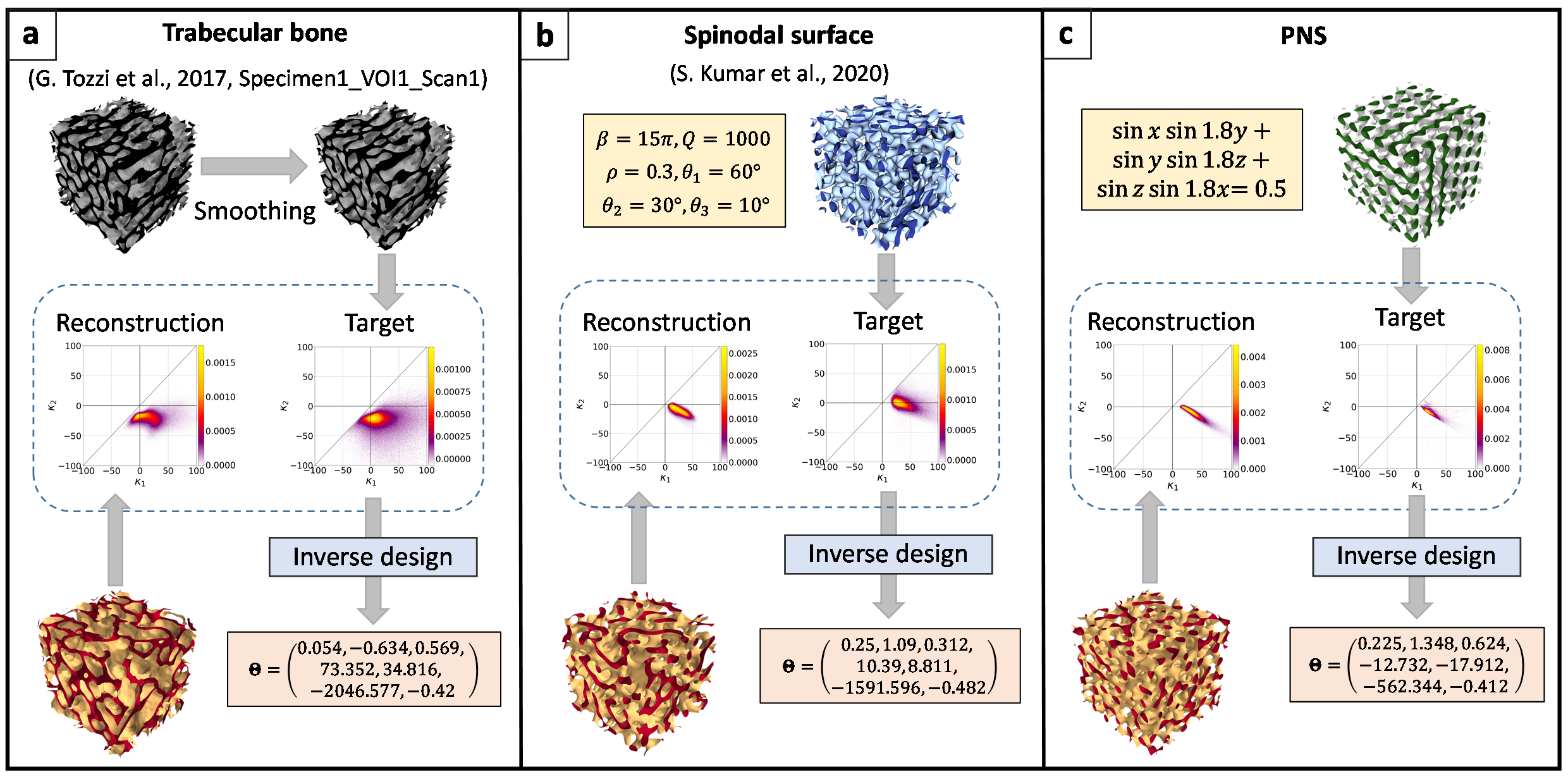}
	\caption{Curvature based surface inverse design applications including (a) a CT scan of trabecular bone specimen taken from Tozzi et al. \cite{tozzi2017}, (b) spinodal surface created from 6 design parameters from Kumar et al. \cite{kumar2020} and (c) PNS surface governed by an implicit equation. The curvature profile of targeted structures are queried through the \textit{i}-NN which outputs the design parameters $\bfTheta$. The corresponding topology is then generated through the phase-field approach and the curvature profile is reconstructed for comparison with the original target. Additional benchmarks are presented in the Supporting Information Section S5.}
	\label{fig:application}
\end{figure}

To demonstrate the generalization ability of our approach, we inverse design the topologies for   tailored curvature profiles that are representative of three different (meta-)material/structural classes.
\begin{enumerate}[label=(\roman*)]
	% ---------------------------------------------
	% ---------------------------------------------
	% ---------------------------------------------
	\item \textbf{Benchmark 1: Trabecular bone.} Recent works \cite{CALLENS2020, Yang2022}  have shown the important role of substrate curvature in growth of bone cells on additively manufactured implants. Motivated by this, we target the microstructural topology of bone to benchmark our inverse design framework. Specifically, we consider a trabecular bone sample from Tozzi et al. \cite{tozzi2017} (see \figurename~\ref{fig:application}a). The microstructural data is available as grayscale 3D-voxelated image obtained directly from micro-computed tomography. The surface representation of the trabecular bone sample is extracted by using an appropriate image processing and smoothing algorithm, the details of which are provided in Supporting Information Section S4. 
	% ---------------------------------------------
	% ---------------------------------------------
	% ---------------------------------------------
	\item \textbf{Benchmark 2: Spinodal decomposition.} Diffusion-driven spinodal decomposition in a binary phase system can produce complex and diverse topologies (which are also used in spinodoid metamaterials  \cite{kumar2020}). Here, we briefly review the formulation of topologies emergent from spinodal decomposition and refer to refs. \cite{kumar2020,zheng2021data} for  details. The early stage of a spinodal decomposition process is mathematically described by a Gaussian Random Field (GRF), i.e., a linear superposition of $Q\gg 1$ standing waves:
	\be
	\varphi(\bfx) = \sqrt{\frac{2}{Q}}\sum_{q=1}^Q\cos(\beta\bfv_q \cdot \bfx + \gamma_q), 
	\ee
	where $\varphi:\Omega\rightarrow\Rset$ and $\beta>0$ denote a phase field and a constant wavenumber, respectively; the latter determines the microstructural lengthscale of the former. $\bfv_q$ and $\gamma_q$ denote, respectively, the wave vector and phase angle of $q^\text{th}$ standing wave and are randomly and uniformly sampled as
	\be
	\begin{aligned}
		\bfv_q&\sim \calU\left(
		\left\{
		\bfv \in \text{S}^2 : \left(\left|\bfv\cdot\hat\bfe_1\right|>\cos\theta_1\right)\oplus
		\left(\left|\bfv\cdot\hat\bfe_2\right|>\cos\theta_2\right)\oplus
		\left(\left|\bfv\cdot\hat\bfe_3\right|>\cos\theta_3\right)
		\right\}
		\right),\\
		\gamma_p &\sim \calU([0,2\pi)),
	\end{aligned}
	\ee
	where $\{\hat\bfe_1,\hat\bfe_2,\hat\bfe_3\}$ denote $\text{S}^2$ is the set of all unit vectors. Parameters $\theta_1,\theta_2,\theta_3\in[0,\pi/2)$ restrict  the directional distribution of the wave vectors to respective angles from the Cartesian basis vectors. This parametric distribution is interpreted as anisotropic mobility along preferred directions during the spinodal decomposition process. The final topology is obtained by extracting the phase interface which is given by a level set as
	\be
	\varphi = \sqrt{2}\text{erf}^{-1}(2\rho-1),
	\ee
	where $\rho\in[0,1]$ is the volume fraction of one of the phases. For the purpose of this benchmark, we generate a spinodoid topology (see \figurename~\ref{fig:application}b) with $\beta=15\pi$, $Q=1000$, $\theta_1=60^\circ$, $\theta_2=30^\circ$, $\theta_3=10^\circ$, and $\rho=0.3$.
	% ---------------------------------------------
	% ---------------------------------------------
	% ---------------------------------------------
	\item \textbf{Benchmark 3: Periodic nodal surfaces (PNS).} These are smooth implicit surfaces composed of Fourier series components \cite{leoni1998,GANDY2001} and are widely studied across different fields in mathematics,\cite{karcher1996,meeks1990,YANG2010} chemistry, \cite{nesper2001,baena2021,leoni1998} mechanics,\cite{kapfer2011,bonatti2019} and more. PNS are particularly popular as closed-form approximations of minimal surfaces (i.e., zero mean curvature) such as the famous Schwarz \cite{ross1992,guo2022} and Schoen \cite{ciliberto2015,rito2016} surfaces. Since minimal surfaces present a curvature distribution  ($\kappa_2=-\kappa_1$) trivial for a benchmark, here we consider a non-minimal PNS given by
	\be
	\sin{(x)}\sin{(1.8 y)}+\sin{(y)}\sin{(1.8 z)}+\sin{(z)}\sin{(1.8 x)} = 0.5
	\ee
	and visualized in \figurename~\ref{fig:application}c.
\end{enumerate}

Across all the three benchmarks, we scale the topologies to the same domain size as $\Omega$ and extract the curvature profile which is passed as a design target to the \textit{i}-NN. For each benchmark, the \textit{i}-NN successfully predicts a topology with a curvature profile that closely matches the target (see  \figurename~\ref{fig:application} and Supporting Information Section S5). Remarkably, the \textit{i}-NN achieves this despite no prior information about trabecular bone, spinodal decomposition, or PNS topologies during the training stage and indicates excellent generalization well beyond the scope of the training dataset. The \textit{i}-NN is also able to design for similar curvature profiles despite key geometrical differences such as periodicity (in the PNS benchmark) and anisotropy (in all three benchmarks)  in the target, which highlights further the ill-posedness of the inverse design challenge and the benefit of the proposed approach.  In addition, the generalization capability is particularly useful for bio-mimetic implants and scaffolds where large training datasets (of e.g., bone) are particularly scarce, and small amounts of patient-/site-specific data can be used for fine-tuning or transfer learning of the pre-trained \textit{i}-NN for improved accuracy.

\subsection{Curvature determines mechanics}
For slender structures -- both natural (e.g., skull, egg shells, sea shells) and man-made (e.g., aircraft fuselage, pressure vessels, domes), curved geometries are preferred as structural elements over flat ones due to their ability to redistribute applied loads into the stronger in-plane stretching-dominated behavior as opposed to the weaker out-of-plane bending-dominated behavior \cite{portela2020extreme}. In the light of microstructures with tailorable curvature profiles, we explore this interplay between curvature and mechanics across our design space. 

\begin{figure}
	\centering
	\includegraphics[width=\textwidth]{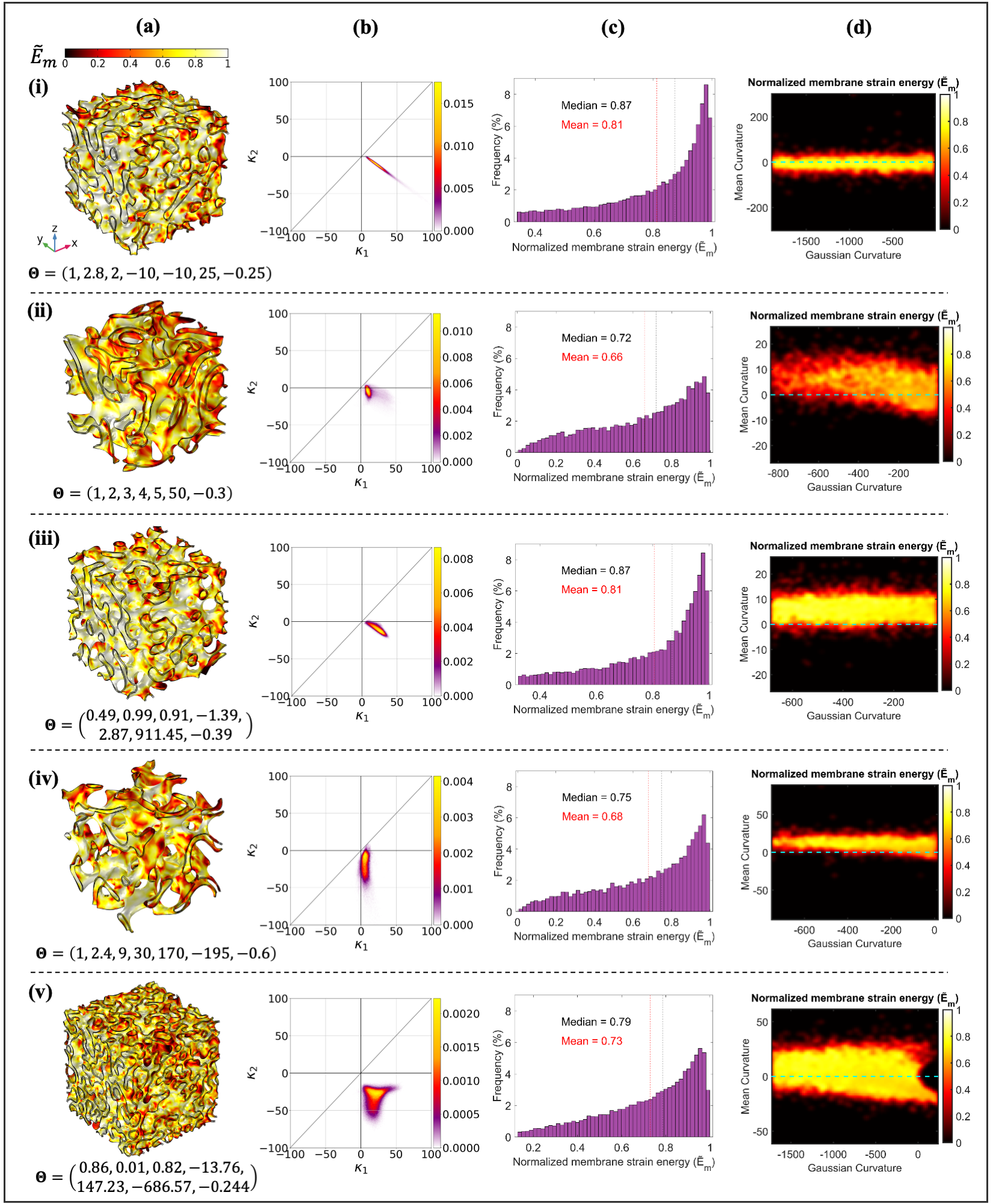}
	\captionsetup{justification=justified}
	\caption{
		Interplay of membrane and bending strain energies in diverse inverse-designed shellular topologies (numbered (i)-(v)) under unit tensile load. \textbf{(a)} Distribution of normalized membrane strain energy $\tilde E_m$ in the topologies and \textbf{(b)} their corresponding curvature profiles.  \textbf{(c)} Distribution of $\tilde E_m$ over the surface area of each topology. Topologies with close-to-zero mean curvature ((i) \& (iii)) show relatively higher values of $\tilde E_m$, indicating a stretching-dominated behavior. \textbf{(d)} Distribution of mean and Gaussian curvatures; colored by the mean of $\tilde E_m$ for the corresponding curvature values found within a topology. Higher $\tilde E_m$ (i.e., stretching-dominated deformation) is observed in the proximity of zero mean curvature (dashed cyan line) compared to other curvatures.}
	\label{fig:mechanics}
\end{figure}

Specifically, we consider five representative topologies with diverse curvature profiles (see \figurename~\ref{fig:mechanics}) obtained from the inverse design strategy. Each topology is modeled as a shellular structure with the shell thickness equal to $1\%$ of the domain length (recall $\Omega=[0,100]^3$) and made of a linearly elastic material with Poisson's ratio of 0.3 and a unit Young's modulus. Note that we restrict ourselves to curvature distributions that admit bicontinuous topologies only as it would be physically required for a self-standing shellular structure. We perform finite element analysis of the shellular structure undergoing quasi-static uniaxial tension, using six degrees of freedom quadratic shell elements with the \textit{Mixed Interpolation of Tensorial Components} formulation (MITC).\cite{bathe1986formulation} The choice of Young's modulus and applied load are arbitrary as they do not affect the following analysis. 

For each case, we investigate the distribution of the total in-plane membrane strain energy $E_m$ and out-of-plane bending strain energy $E_b$ (obtained by summing over the membrane and bending strain energies of all the shell elements, respectively). \figurename~\ref{fig:mechanics}a shows the spatial distribution of the normalized membrane strain energy, i.e., $\tilde E_m = E_m/(E_m+E_b)$, with $\tilde E_m=1$ and $\tilde E_m=0$ denoting  pure stretching- and bending-dominated deformations respectively.
In \figurenames~\ref{fig:mechanics}b and \ref{fig:mechanics}c, the topologies with curvature distribution closer to zero mean curvature (first and third row) show higher mean and median in the distribution of $\tilde E_m$, relative to the others, indicating higher stretching-dominated deformation. \figurenames~\ref{fig:mechanics}d  further shows the distribution of element-wise $\tilde E_m$ with respect to the mean and Gaussian curvature. Notably, elements with or close to zero mean curvature predominantly store stretching-dominated energy, which largely corroborates the beneficial mechanical properties of TPMS observed across several prior works.\cite{Shevchenko2023,MASKERY2018,QIU2023}. Moreover, a similar study is conducted under a shear load in Supporting Information Section S5, which shows that the distribution of strain energy between membrane and bending modes follows the same correlation with curvature profile under varying loading conditions.

\section{Conclusion}

The presented framework enables inverse design of smooth topologies with tailored curvature profiles. This is achieved by training a dual deep-NN setup to surrogate a phase field driven by a curvature-based energy functional and then invert the process. Our approach provides excellent generalization beyond both the design space as well as the space of the training data -- as evidenced by successful reconstruction of curvature of topologies from different data sources, including microtomography of trabecular bone, spinodal surfaces, and implicit nodal surfaces. Notably, the ML framework achieves accurate reconstructions without any prior knowledge about these structures and their symmetries during the learning phase. Linking with mechanics, the analysis of strain energy distribution reveals a correlation between curvature profile and stretching vs.~bending dominated deformations, which can be exploited for tailoring the topology for improved mechanical resilience. Overall, our framework unlocks curvature as a new design modality with applications in mechanical metamaterials and bio-scaffolds/implants.

Future directions of interest may include incorporating anisotropic and higher-order curvature descriptors for a more generalizable design space and integrating topology curvature with nonlinear mechanical response in inverse design. Specifically, two surfaces can have similar curvature distributions, yet the orientation distribution of those surfaces can be significantly different. Higher-order curvature descriptors such as Minkowski functionals \cite{SchrderTurk2011} can be introduced in the energetics of the phase-field methods to preferentially bias both surface curvature and orientation distributions simultaneously. This will expand the design space to include anisotropic designs, and can be relevant for controllable anisotropy in mechanical response of metamaterials as well as mimicking topological anisotropy in bio-scaffolds for improved bio-compatibility. Additionally, the incorporation of spatially-varying design parameters will facilitate the creation of topologies with spatially graded curvature variation (commonly observed in biological tissues and bones), offering the potential for a more comprehensive modeling approach. 

\bigskip
\bigskip
\bigskip
\bigskip

\raggedright
\textbf{Supporting Information} \par %Please delete the Suppporting Information statement if it is not applicable. Please supply Supporting Information in another file. Supporting information should not be provided in .tex format

\noindent
{Supporting Information is available from the Wiley Online Library or from the author.}

% Acknowledgements
\medskip
\raggedright
\textbf{Acknowledgements} \par 

\noindent
{Y. Guo gratefully acknowledges financial support from the China Scholarship Council (No.202006260125).}

\medskip
\raggedright
\textbf{Conflict of Interest} \par %delete if not applicable))
The authors declare no conflict of interest.

\medskip
\raggedright
\textbf{Author contributions} \par

\noindent
{Y.G. developed and implemented the inverse design algorithm and performed numerical experiments. S.S. performed the mechanics simulations and analyzed the results. S.K. conceived the research. S.S. and S.K. supervised the research. All authors wrote and approved the final manuscript.}

\medskip
\raggedright
\textbf{Code Availability Statement} \par 

\noindent
{The codes developed during the current study are available at \href{https://github.com/mmc-group/inverse-designed-surface-curvatures}{https://github.com/mmc-group/inverse-designed-surface-curvatures}.}

\medskip
\raggedright
\textbf{Data Availability Statement} \par %delete if not applicable))

\noindent
{The datasets generated during the current study are available at \href{https://github.com/mmc-group/inverse-designed-surface-curvatures}{https://github.com/mmc-group/inverse-designed-surface-curvatures}.}

% References
\medskip

%\bibliographystyle{MSP}
%
%\bibliography{Bib}

\bibliography{Bib}

\begin{thebibliography}{10}
\urlstyle{rm}
\expandafter\ifx\csname url\endcsname\relax
  \def\url#1{\texttt{#1}}\fi
\expandafter\ifx\csname urlprefix\endcsname\relax\def\urlprefix{URL: }\fi
\expandafter\ifx\csname doiprefix\endcsname\relax\def\doiprefix{DOI: }\fi
\providecommand{\bibinfo}[2]{#2}
\providecommand{\eprint}[2][]{\url{#2}}

\bibitem{taylor1976}
\bibinfo{author}{Taylor, J.~E.}
\newblock \bibinfo{journal}{\bibinfo{title}{The structure of singularities in
  soap-bubble-like and soap-film-like minimal surfaces}}.
\newblock {\emph{\JournalTitle{Annals of Mathematics}}}
  \textbf{\bibinfo{volume}{103}}, \bibinfo{pages}{489--539}
  (\bibinfo{year}{1976}).

\bibitem{almgren1976}
\bibinfo{author}{Almgren, F.~J.} \& \bibinfo{author}{Taylor, J.~E.}
\newblock \bibinfo{journal}{\bibinfo{title}{The geometry of soap films and soap
  bubbles}}.
\newblock {\emph{\JournalTitle{Scientific American}}}
  \textbf{\bibinfo{volume}{235}}, \bibinfo{pages}{82--93}
  (\bibinfo{year}{1976}).

\bibitem{Turing1952}
\bibinfo{author}{{Turing}, A.~M.}
\newblock \bibinfo{journal}{\bibinfo{title}{{The Chemical Basis of
  Morphogenesis}}}.
\newblock {\emph{\JournalTitle{Philosophical Transactions of the Royal Society
  of London Series B}}} \textbf{\bibinfo{volume}{237}},
  \bibinfo{pages}{37--72}, \doiprefix\url{10.1098/rstb.1952.0012}
  (\bibinfo{year}{1952}).

\bibitem{Kondo2017}
\bibinfo{author}{Kondo, S.}
\newblock \bibinfo{journal}{\bibinfo{title}{An updated kernel-based turing
  model for studying the mechanisms of biological pattern formation}}.
\newblock {\emph{\JournalTitle{Journal of Theoretical Biology}}}
  \textbf{\bibinfo{volume}{414}}, \bibinfo{pages}{120--127},
  \doiprefix\url{10.1016/j.jtbi.2016.11.003} (\bibinfo{year}{2017}).

\bibitem{callens2021}
\bibinfo{author}{Callens, S.~J.}, \bibinfo{author}{{Tourolle né Betts},
  D.~C.}, \bibinfo{author}{Müller, R.} \& \bibinfo{author}{Zadpoor, A.~A.}
\newblock \bibinfo{journal}{\bibinfo{title}{The local and global geometry of
  trabecular bone}}.
\newblock {\emph{\JournalTitle{Acta Biomaterialia}}}
  \textbf{\bibinfo{volume}{130}}, \bibinfo{pages}{343--361},
  \doiprefix\url{https://doi.org/10.1016/j.actbio.2021.06.013}
  (\bibinfo{year}{2021}).

\bibitem{Ralph2009}
\bibinfo{author}{Ralph, M.}
\newblock \bibinfo{journal}{\bibinfo{title}{Hierarchical microimaging of bone
  structure and function.}}
\newblock {\emph{\JournalTitle{Nature reviews. Rheumatology}}}
  \textbf{\bibinfo{volume}{5}}, \bibinfo{pages}{373--81},
  \doiprefix\url{10.1038/nrrheum.2009.107} (\bibinfo{year}{2009}).

\bibitem{Puhka2012}
\bibinfo{author}{Puhka, M.}, \bibinfo{author}{Joensuu, M.},
  \bibinfo{author}{Vihinen, H.}, \bibinfo{author}{Belevich, I.} \&
  \bibinfo{author}{Jokitalo, E.}
\newblock \bibinfo{journal}{\bibinfo{title}{Progressive sheet-to-tubule
  transformation is a general mechanism for endoplasmic reticulum partitioning
  in dividing mammalian cells}}.
\newblock {\emph{\JournalTitle{Molecular Biology of the Cell}}}
  \textbf{\bibinfo{volume}{23}}, \bibinfo{pages}{2424--2432},
  \doiprefix\url{10.1091/mbc.e10-12-0950} (\bibinfo{year}{2012}).

\bibitem{ramasamy2017}
\bibinfo{author}{Ramasamy, S.~K.}
\newblock \bibinfo{journal}{\bibinfo{title}{Structure and functions of blood
  vessels and vascular niches in bone}}.
\newblock {\emph{\JournalTitle{Stem cells international}}}
  \textbf{\bibinfo{volume}{2017}} (\bibinfo{year}{2017}).

\bibitem{sivaraj2016}
\bibinfo{author}{Sivaraj, K.~K.} \& \bibinfo{author}{Adams, R.~H.}
\newblock \bibinfo{journal}{\bibinfo{title}{Blood vessel formation and function
  in bone}}.
\newblock {\emph{\JournalTitle{Development}}} \textbf{\bibinfo{volume}{143}},
  \bibinfo{pages}{2706--2715} (\bibinfo{year}{2016}).

\bibitem{Erlebacher2001}
\bibinfo{author}{Erlebacher, J.}, \bibinfo{author}{Aziz, M.~J.},
  \bibinfo{author}{Karma, A.}, \bibinfo{author}{Dimitrov, N.} \&
  \bibinfo{author}{Sieradzki, K.}
\newblock \bibinfo{journal}{\bibinfo{title}{Evolution of nanoporosity in
  dealloying}}.
\newblock {\emph{\JournalTitle{Nature}}} \textbf{\bibinfo{volume}{410}},
  \bibinfo{pages}{450--453}, \doiprefix\url{10.1038/35068529}
  (\bibinfo{year}{2001}).

\bibitem{portela2020extreme}
\bibinfo{author}{Portela, C.~M.} \emph{et~al.}
\newblock \bibinfo{journal}{\bibinfo{title}{Extreme mechanical resilience of
  self-assembled nanolabyrinthine materials}}.
\newblock {\emph{\JournalTitle{Proceedings of the National Academy of
  Sciences}}} \textbf{\bibinfo{volume}{117}}, \bibinfo{pages}{5686--5693}
  (\bibinfo{year}{2020}).

\bibitem{hill2019}
\bibinfo{author}{Hill, J.~D.} \& \bibinfo{author}{Millett, P.~C.}
\newblock \bibinfo{journal}{\bibinfo{title}{Directed self-assembly in diblock
  copolymer thin films for uniform hemisphere pattern formation}}.
\newblock {\emph{\JournalTitle{Macromolecules}}} \textbf{\bibinfo{volume}{52}},
  \bibinfo{pages}{9495--9503}, \doiprefix\url{10.1021/acs.macromol.9b01545}
  (\bibinfo{year}{2019}).

\bibitem{jung2019}
\bibinfo{author}{Jung, D.~S.} \emph{et~al.}
\newblock \bibinfo{journal}{\bibinfo{title}{Pattern formation of metal–oxide
  hybrid nanostructures via the self-assembly of di-block copolymer blends}}.
\newblock {\emph{\JournalTitle{Nanoscale}}} \textbf{\bibinfo{volume}{11}},
  \bibinfo{pages}{18559--18567}, \doiprefix\url{10.1039/C9NR04038B}
  (\bibinfo{year}{2019}).

\bibitem{ren2018}
\bibinfo{author}{Ren, J.} \emph{et~al.}
\newblock \bibinfo{journal}{\bibinfo{title}{Engineering the kinetics of
  directed self-assembly of block copolymers toward fast and defect-free
  assembly}}.
\newblock {\emph{\JournalTitle{ACS Applied Materials \& Interfaces}}}
  \textbf{\bibinfo{volume}{10}}, \bibinfo{pages}{23414--23423},
  \doiprefix\url{10.1021/acsami.8b05247} (\bibinfo{year}{2018}).

\bibitem{paul2021}
\bibinfo{author}{Castillo, P.} \& \bibinfo{author}{Gómez, S.}
\newblock \bibinfo{journal}{\bibinfo{title}{An interpolatory directional
  splitting-local discontinuous galerkin method with application to pattern
  formation in 2d/3d}}.
\newblock {\emph{\JournalTitle{Applied Mathematics and Computation}}}
  \textbf{\bibinfo{volume}{397}}, \bibinfo{pages}{125984},
  \doiprefix\url{https://doi.org/10.1016/j.amc.2021.125984}
  (\bibinfo{year}{2021}).

\bibitem{jiang2021}
\bibinfo{author}{Jiang, M.}, \bibinfo{author}{Zhang, J.}, \bibinfo{author}{Zhu,
  J.}, \bibinfo{author}{Yu, X.} \& \bibinfo{author}{Bevilacqua, L.}
\newblock \bibinfo{journal}{\bibinfo{title}{Numerical simulation for clustering
  and pattern formation in active colloids with mass-preserving characteristic
  finite element method}}.
\newblock {\emph{\JournalTitle{Computer Methods in Applied Mechanics and
  Engineering}}} \textbf{\bibinfo{volume}{381}}, \bibinfo{pages}{113806},
  \doiprefix\url{https://doi.org/10.1016/j.cma.2021.113806}
  (\bibinfo{year}{2021}).

\bibitem{anzaki2021}
\bibinfo{author}{Anzaki, R.} \emph{et~al.}
\newblock \bibinfo{journal}{\bibinfo{title}{Phase prediction method for pattern
  formation in time-dependent ginzburg-landau dynamics for kinetic ising model
  without a priori assumptions of domain patterns}}.
\newblock {\emph{\JournalTitle{Phys. Rev. B}}} \textbf{\bibinfo{volume}{103}},
  \bibinfo{pages}{094408}, \doiprefix\url{10.1103/PhysRevB.103.094408}
  (\bibinfo{year}{2021}).

\bibitem{diaz2022}
\bibinfo{author}{Diaz, J.}, \bibinfo{author}{Pinna, M.},
  \bibinfo{author}{Zvelindovsky, A.~V.} \& \bibinfo{author}{Pagonabarraga, I.}
\newblock \bibinfo{journal}{\bibinfo{title}{Nematic ordering of anisotropic
  nanoparticles in block copolymers}}.
\newblock {\emph{\JournalTitle{Advanced Theory and Simulations}}}
  \textbf{\bibinfo{volume}{5}}, \bibinfo{pages}{2100433},
  \doiprefix\url{https://doi.org/10.1002/adts.202100433}
  (\bibinfo{year}{2022}).

\bibitem{ankudinov2022}
\bibinfo{author}{Ankudinov, V.} \& \bibinfo{author}{Galenko, P.~K.}
\newblock \bibinfo{journal}{\bibinfo{title}{Structure diagram and dynamics of
  formation of hexagonal boron nitride in phase-field crystal model}}.
\newblock {\emph{\JournalTitle{Philosophical Transactions of the Royal Society
  A: Mathematical, Physical and Engineering Sciences}}}
  \textbf{\bibinfo{volume}{380}}, \bibinfo{pages}{20200318},
  \doiprefix\url{10.1098/rsta.2020.0318} (\bibinfo{year}{2022}).

\bibitem{bone}
\bibinfo{author}{{Laboratoires Servier}}.
\newblock \bibinfo{title}{{Cancellous bone}}.
\newblock \bibinfo{howpublished}{Wikimedia Commons} (\bibinfo{year}{2019}).
\newblock
  \urlprefix\url{https://commons.wikimedia.org/wiki/File:Spongy\_bone\_-\_Trabecular\_bone\_-\_Normal\_trabecular\_bone\_--\_Smart-Servier.png}.

\bibitem{soapfilm}
\bibinfo{author}{{Blinking Spirit}}.
\newblock \bibinfo{title}{{Photo of a soap bubble creating a catenoid}}.
\newblock \bibinfo{howpublished}{Wikimedia Commons} (\bibinfo{year}{2006}).
\newblock
  \urlprefix\url{https://commons.wikimedia.org/wiki/File:Bulle\_cat%C3%A9no%C3%AFde.png}.

\bibitem{ceramic}
\bibinfo{author}{{Onnovisser1979}}.
\newblock \bibinfo{title}{{Photo of porous ceramic granule, photo shot by
  Michel Porro}}.
\newblock \bibinfo{howpublished}{Wikimedia Commons} (\bibinfo{year}{2012}).
\newblock
  \urlprefix\url{https://commons.wikimedia.org/wiki/File:Cam\_Bioceramics\_Large\_Porous\_Granule.png}.

\bibitem{vessels}
\bibinfo{author}{{I'm in the garden}}.
\newblock \bibinfo{title}{{Injection preparation; Universum Bremen}}.
\newblock \bibinfo{howpublished}{Wikimedia Commons} (\bibinfo{year}{2009}).
\newblock
  \urlprefix\url{https://commons.wikimedia.org/wiki/File:Blutgef%C3%A4%C3%9Fe\_0387.JPG}.

\bibitem{Soyarslan2018}
\bibinfo{author}{Soyarslan, C.}, \bibinfo{author}{Bargmann, S.},
  \bibinfo{author}{Pradas, M.} \& \bibinfo{author}{Weissm{\"u}ller, J.}
\newblock \bibinfo{journal}{\bibinfo{title}{3d stochastic bicontinuous
  microstructures: Generation, topology and elasticity}}.
\newblock {\emph{\JournalTitle{Acta materialia}}}
  \textbf{\bibinfo{volume}{149}}, \bibinfo{pages}{326--340}
  (\bibinfo{year}{2018}).

\bibitem{Meng2019}
\bibinfo{author}{Hsieh, M.-T.}, \bibinfo{author}{Endo, B.},
  \bibinfo{author}{Zhang, Y.}, \bibinfo{author}{Bauer, J.} \&
  \bibinfo{author}{Valdevit, L.}
\newblock \bibinfo{journal}{\bibinfo{title}{The mechanical response of cellular
  materials with spinodal topologies}}.
\newblock {\emph{\JournalTitle{Journal of the Mechanics and Physics of
  Solids}}} \textbf{\bibinfo{volume}{125}}, \bibinfo{pages}{401--419},
  \doiprefix\url{https://doi.org/10.1016/j.jmps.2019.01.002}
  (\bibinfo{year}{2019}).

\bibitem{kumar2020}
\bibinfo{author}{Kumar, S.}, \bibinfo{author}{Tan, S.}, \bibinfo{author}{Zheng,
  L.} \& \bibinfo{author}{Kochmann, D.~M.}
\newblock \bibinfo{journal}{\bibinfo{title}{Inverse-designed spinodoid
  metamaterials}}.
\newblock {\emph{\JournalTitle{npj Computational Materials}}}
  \textbf{\bibinfo{volume}{6}}, \bibinfo{pages}{1--10},
  \doiprefix\url{https://doi.org/10.1038/s41524-020-0341-6}
  (\bibinfo{year}{2020}).

\bibitem{Meng2021}
\bibinfo{author}{Hsieh, M.-T.}, \bibinfo{author}{Begley, M.~R.} \&
  \bibinfo{author}{Valdevit, L.}
\newblock \bibinfo{journal}{\bibinfo{title}{Architected implant designs for
  long bones: Advantages of minimal surface-based topologies}}.
\newblock {\emph{\JournalTitle{Materials \& Design}}}
  \textbf{\bibinfo{volume}{207}}, \bibinfo{pages}{109838},
  \doiprefix\url{https://doi.org/10.1016/j.matdes.2021.109838}
  (\bibinfo{year}{2021}).

\bibitem{zheng2021data}
\bibinfo{author}{Zheng, L.}, \bibinfo{author}{Kumar, S.} \&
  \bibinfo{author}{Kochmann, D.~M.}
\newblock \bibinfo{journal}{\bibinfo{title}{Data-driven topology optimization
  of spinodoid metamaterials with seamlessly tunable anisotropy}}.
\newblock {\emph{\JournalTitle{Computer Methods in Applied Mechanics and
  Engineering}}} \textbf{\bibinfo{volume}{383}}, \bibinfo{pages}{113894}
  (\bibinfo{year}{2021}).

\bibitem{GuellIzard2019}
\bibinfo{author}{Izard, A.~G.}, \bibinfo{author}{Bauer, J.},
  \bibinfo{author}{Crook, C.}, \bibinfo{author}{Turlo, V.} \&
  \bibinfo{author}{Valdevit, L.}
\newblock \bibinfo{journal}{\bibinfo{title}{Ultrahigh energy absorption
  multifunctional spinodal nanoarchitectures}}.
\newblock {\emph{\JournalTitle{Small}}} \textbf{\bibinfo{volume}{15}},
  \bibinfo{pages}{1903834}, \doiprefix\url{10.1002/smll.201903834}
  (\bibinfo{year}{2019}).

\bibitem{Roding2022}
\bibinfo{author}{R\"{o}ding, M.}, \bibinfo{author}{Sk\"{a}rstr\"{o}m, V.~W.} \&
  \bibinfo{author}{Lor{\'{e}}n, N.}
\newblock \bibinfo{journal}{\bibinfo{title}{Inverse design of anisotropic
  spinodoid materials with prescribed diffusivity}}.
\newblock {\emph{\JournalTitle{Scientific Reports}}}
  \textbf{\bibinfo{volume}{12}}, \doiprefix\url{10.1038/s41598-022-21451-6}
  (\bibinfo{year}{2022}).

\bibitem{lou2022}
\bibinfo{author}{Lou, Y.}, \bibinfo{author}{Rupprecht, J.-F.},
  \bibinfo{author}{Hiraiwa, T.} \& \bibinfo{author}{Saunders, T.~E.}
\newblock \bibinfo{journal}{\bibinfo{title}{Curvature-induced cell
  rearrangements in biological tissues}}.
\newblock {\emph{\JournalTitle{bioRxiv}}}
  \doiprefix\url{10.1101/2022.05.18.492428} (\bibinfo{year}{2022}).

\bibitem{Werner2016}
\bibinfo{author}{Werner, M.} \emph{et~al.}
\newblock \bibinfo{journal}{\bibinfo{title}{Surface curvature differentially
  regulates stem cell migration and differentiation via altered attachment
  morphology and nuclear deformation}}.
\newblock {\emph{\JournalTitle{Advanced Science}}}
  \textbf{\bibinfo{volume}{4}}, \bibinfo{pages}{1600347},
  \doiprefix\url{10.1002/advs.201600347} (\bibinfo{year}{2016}).

\bibitem{zadpoor2015}
\bibinfo{author}{Zadpoor, A.~A.}
\newblock \bibinfo{journal}{\bibinfo{title}{Bone tissue regeneration: the role
  of scaffold geometry}}.
\newblock {\emph{\JournalTitle{Biomater. Sci.}}} \textbf{\bibinfo{volume}{3}},
  \bibinfo{pages}{231--245}, \doiprefix\url{10.1039/C4BM00291A}
  (\bibinfo{year}{2015}).

\bibitem{zhang2022}
\bibinfo{author}{Zhang, Y.} \emph{et~al.}
\newblock \bibinfo{journal}{\bibinfo{title}{In silico and in vivo studies of
  the effect of surface curvature on the osteoconduction of porous scaffolds}}.
\newblock {\emph{\JournalTitle{Biotechnology and Bioengineering}}}
  \textbf{\bibinfo{volume}{119}}, \bibinfo{pages}{591--604},
  \doiprefix\url{https://doi.org/10.1002/bit.27976} (\bibinfo{year}{2022}).

\bibitem{Sebastien2020}
\bibinfo{author}{Callens, S.~J.}, \bibinfo{author}{Uyttendaele, R.~J.},
  \bibinfo{author}{Fratila-Apachitei, L.~E.} \& \bibinfo{author}{Zadpoor,
  A.~A.}
\newblock \bibinfo{journal}{\bibinfo{title}{Substrate curvature as a cue to
  guide spatiotemporal cell and tissue organization}}.
\newblock {\emph{\JournalTitle{Biomaterials}}} \textbf{\bibinfo{volume}{232}},
  \bibinfo{pages}{119739},
  \doiprefix\url{https://doi.org/10.1016/j.biomaterials.2019.119739}
  (\bibinfo{year}{2020}).

\bibitem{Callens2023}
\bibinfo{author}{Callens, S. J.~P.} \emph{et~al.}
\newblock \bibinfo{journal}{\bibinfo{title}{Emergent collective organization of
  bone cells in complex curvature fields}}.
\newblock {\emph{\JournalTitle{Nature Communications}}}
  \textbf{\bibinfo{volume}{14}}, \doiprefix\url{10.1038/s41467-023-36436-w}
  (\bibinfo{year}{2023}).

\bibitem{Song2021}
\bibinfo{author}{Song, A.}
\newblock \bibinfo{journal}{\bibinfo{title}{Generation of tubular and
  membranous shape textures with curvature functionals}}.
\newblock {\emph{\JournalTitle{Journal of Mathematical Imaging and Vision}}}
  \textbf{\bibinfo{volume}{64}}, \bibinfo{pages}{17--40},
  \doiprefix\url{10.1007/s10851-021-01049-9} (\bibinfo{year}{2021}).

\bibitem{Cahn1958}
\bibinfo{author}{Cahn, J.~W.} \& \bibinfo{author}{Hilliard, J.~E.}
\newblock \bibinfo{journal}{\bibinfo{title}{Free energy of a nonuniform system.
  i. interfacial free energy}}.
\newblock {\emph{\JournalTitle{The Journal of Chemical Physics}}}
  \textbf{\bibinfo{volume}{28}}, \bibinfo{pages}{258--267},
  \doiprefix\url{10.1063/1.1744102} (\bibinfo{year}{1958}).

\bibitem{Sigmund2013}
\bibinfo{author}{Sigmund, O.} \& \bibinfo{author}{Maute, K.}
\newblock \bibinfo{journal}{\bibinfo{title}{Topology optimization approaches}}.
\newblock {\emph{\JournalTitle{Structural and Multidisciplinary Optimization}}}
  \textbf{\bibinfo{volume}{48}}, \bibinfo{pages}{1031--1055},
  \doiprefix\url{https://doi.org/10.1007/s00158-013-0978-6}
  (\bibinfo{year}{2013}).

\bibitem{zhai2024topology}
\bibinfo{author}{Zhai, X.}, \bibinfo{author}{Wang, W.}, \bibinfo{author}{Chen,
  F.} \& \bibinfo{author}{Wu, J.}
\newblock \bibinfo{journal}{\bibinfo{title}{Topology optimization of
  differentiable microstructures}}.
\newblock {\emph{\JournalTitle{Computer Methods in Applied Mechanics and
  Engineering}}} \textbf{\bibinfo{volume}{418}}, \bibinfo{pages}{116530}
  (\bibinfo{year}{2024}).

\bibitem{gao2020comprehensive}
\bibinfo{author}{Gao, J.}, \bibinfo{author}{Xiao, M.}, \bibinfo{author}{Zhang,
  Y.} \& \bibinfo{author}{Gao, L.}
\newblock \bibinfo{journal}{\bibinfo{title}{A comprehensive review of
  isogeometric topology optimization: methods, applications and prospects}}.
\newblock {\emph{\JournalTitle{Chinese Journal of Mechanical Engineering}}}
  \textbf{\bibinfo{volume}{33}}, \bibinfo{pages}{1--14} (\bibinfo{year}{2020}).

\bibitem{gibson2010cellular}
\bibinfo{author}{Gibson, L.~J.}, \bibinfo{author}{Ashby, M.~F.} \&
  \bibinfo{author}{Harley, B.~A.}
\newblock \emph{\bibinfo{title}{Cellular materials in nature and medicine}}
  (\bibinfo{publisher}{Cambridge University Press}, \bibinfo{year}{2010}).

\bibitem{espinosa2011tablet}
\bibinfo{author}{Espinosa, H.~D.} \emph{et~al.}
\newblock \bibinfo{journal}{\bibinfo{title}{Tablet-level origin of toughening
  in abalone shells and translation to synthetic composite materials}}.
\newblock {\emph{\JournalTitle{Nature communications}}}
  \textbf{\bibinfo{volume}{2}}, \bibinfo{pages}{173} (\bibinfo{year}{2011}).

\bibitem{barthelat2007experimental}
\bibinfo{author}{Barthelat, F.} \& \bibinfo{author}{Espinosa, H.}
\newblock \bibinfo{journal}{\bibinfo{title}{An experimental investigation of
  deformation and fracture of nacre--mother of pearl}}.
\newblock {\emph{\JournalTitle{Experimental mechanics}}}
  \textbf{\bibinfo{volume}{47}}, \bibinfo{pages}{311--324}
  (\bibinfo{year}{2007}).

\bibitem{barthelat2011toughness}
\bibinfo{author}{Barthelat, F.} \& \bibinfo{author}{Rabiei, R.}
\newblock \bibinfo{journal}{\bibinfo{title}{Toughness amplification in natural
  composites}}.
\newblock {\emph{\JournalTitle{Journal of the Mechanics and Physics of
  Solids}}} \textbf{\bibinfo{volume}{59}}, \bibinfo{pages}{829--840}
  (\bibinfo{year}{2011}).

\bibitem{yang2022gaussian}
\bibinfo{author}{Yang, Y.} \emph{et~al.}
\newblock \bibinfo{journal}{\bibinfo{title}{Gaussian curvature--driven
  direction of cell fate toward osteogenesis with triply periodic minimal
  surface scaffolds}}.
\newblock {\emph{\JournalTitle{Proceedings of the National Academy of
  Sciences}}} \textbf{\bibinfo{volume}{119}}, \bibinfo{pages}{e2206684119}
  (\bibinfo{year}{2022}).

\bibitem{garnett_bayesoptbook_2023}
\bibinfo{author}{Garnett, R.}
\newblock \emph{\bibinfo{title}{{Bayesian Optimization}}}
  (\bibinfo{publisher}{Cambridge University Press}, \bibinfo{year}{2023}).

\bibitem{PinheiroCinelli2021}
\bibinfo{author}{Cinelli, L.~P.}, \bibinfo{author}{Marins, M.~A.},
  \bibinfo{author}{da~Silva, E. A.~B.} \& \bibinfo{author}{Netto, S.~L.}
\newblock \bibinfo{title}{Variational autoencoder}.
\newblock In \emph{\bibinfo{booktitle}{Variational Methods for Machine Learning
  with Applications to Deep Networks}}, \bibinfo{pages}{111--149},
  \doiprefix\url{10.1007/978-3-030-70679-1_5} (\bibinfo{publisher}{Springer
  International Publishing}, \bibinfo{year}{2021}).

\bibitem{Goodfellow2014}
\bibinfo{author}{Goodfellow, I.} \emph{et~al.}
\newblock \bibinfo{title}{Generative adversarial nets}.
\newblock In \bibinfo{editor}{Ghahramani, Z.}, \bibinfo{editor}{Welling, M.},
  \bibinfo{editor}{Cortes, C.}, \bibinfo{editor}{Lawrence, N.} \&
  \bibinfo{editor}{Weinberger, K.} (eds.) \emph{\bibinfo{booktitle}{Advances in
  Neural Information Processing Systems}}, vol.~\bibinfo{volume}{27}
  (\bibinfo{publisher}{Curran Associates, Inc.}, \bibinfo{year}{2014}).

\bibitem{yang2023diffusion}
\bibinfo{author}{Yang, L.} \emph{et~al.}
\newblock \bibinfo{title}{Diffusion models: A comprehensive survey of methods
  and applications} (\bibinfo{year}{2023}).
\newblock \eprint{2209.00796}.

\bibitem{Bastek2021}
\bibinfo{author}{Bastek, J.-H.}, \bibinfo{author}{Kumar, S.},
  \bibinfo{author}{Telgen, B.}, \bibinfo{author}{Glaesener, R.~N.} \&
  \bibinfo{author}{Kochmann, D.~M.}
\newblock \bibinfo{journal}{\bibinfo{title}{Inverting the
  structure{\textendash}property map of truss metamaterials by deep learning}}.
\newblock {\emph{\JournalTitle{Proceedings of the National Academy of
  Sciences}}} \textbf{\bibinfo{volume}{119}},
  \doiprefix\url{10.1073/pnas.2111505119} (\bibinfo{year}{2021}).

\bibitem{Wang2020}
\bibinfo{author}{Wang, L.} \emph{et~al.}
\newblock \bibinfo{journal}{\bibinfo{title}{Deep generative modeling for
  mechanistic-based learning and design of metamaterial systems}}.
\newblock {\emph{\JournalTitle{Computer Methods in Applied Mechanics and
  Engineering}}} \textbf{\bibinfo{volume}{372}}, \bibinfo{pages}{113377},
  \doiprefix\url{10.1016/j.cma.2020.113377} (\bibinfo{year}{2020}).

\bibitem{Bessa2019}
\bibinfo{author}{Bessa, M.~A.}, \bibinfo{author}{Glowacki, P.} \&
  \bibinfo{author}{Houlder, M.}
\newblock \bibinfo{journal}{\bibinfo{title}{Bayesian machine learning in
  metamaterial design: Fragile becomes supercompressible}}.
\newblock {\emph{\JournalTitle{Advanced Materials}}}
  \textbf{\bibinfo{volume}{31}}, \bibinfo{pages}{1904845},
  \doiprefix\url{https://doi.org/10.1002/adma.201904845}
  (\bibinfo{year}{2019}).

\bibitem{vlassis2023denoising}
\bibinfo{author}{Vlassis, N.~N.} \& \bibinfo{author}{Sun, W.}
\newblock \bibinfo{title}{Denoising diffusion algorithm for inverse design of
  microstructures with fine-tuned nonlinear material properties}
  (\bibinfo{year}{2023}).
\newblock \eprint{2302.12881}.

\bibitem{Mao2020}
\bibinfo{author}{Mao, Y.}, \bibinfo{author}{He, Q.} \& \bibinfo{author}{Zhao,
  X.}
\newblock \bibinfo{journal}{\bibinfo{title}{Designing complex architectured
  materials with generative adversarial networks}}.
\newblock {\emph{\JournalTitle{Science Advances}}}
  \textbf{\bibinfo{volume}{6}}, \doiprefix\url{10.1126/sciadv.aaz4169}
  (\bibinfo{year}{2020}).

\bibitem{Gu2018}
\bibinfo{author}{Gu, G.~X.}, \bibinfo{author}{Chen, C.-T.} \&
  \bibinfo{author}{Buehler, M.~J.}
\newblock \bibinfo{journal}{\bibinfo{title}{De novo composite design based on
  machine learning algorithm}}.
\newblock {\emph{\JournalTitle{Extreme Mechanics Letters}}}
  \textbf{\bibinfo{volume}{18}}, \bibinfo{pages}{19--28},
  \doiprefix\url{10.1016/j.eml.2017.10.001} (\bibinfo{year}{2018}).

\bibitem{Malkiel2018}
\bibinfo{author}{Malkiel, I.} \emph{et~al.}
\newblock \bibinfo{journal}{\bibinfo{title}{Plasmonic nanostructure design and
  characterization via deep learning}}.
\newblock {\emph{\JournalTitle{Light: Science {\&} Applications}}}
  \textbf{\bibinfo{volume}{7}}, \doiprefix\url{10.1038/s41377-018-0060-7}
  (\bibinfo{year}{2018}).

\bibitem{Suwardi2021}
\bibinfo{author}{Suwardi, A.} \emph{et~al.}
\newblock \bibinfo{journal}{\bibinfo{title}{Machine learning-driven
  biomaterials evolution}}.
\newblock {\emph{\JournalTitle{Advanced Materials}}}
  \textbf{\bibinfo{volume}{34}}, \bibinfo{pages}{2102703},
  \doiprefix\url{10.1002/adma.202102703} (\bibinfo{year}{2021}).

\bibitem{VANTSANT2023}
\bibinfo{author}{{Van ’t Sant}, S.}, \bibinfo{author}{Thakolkaran, P.},
  \bibinfo{author}{Martínez, J.} \& \bibinfo{author}{Kumar, S.}
\newblock \bibinfo{journal}{\bibinfo{title}{Inverse-designed growth-based
  cellular metamaterials}}.
\newblock {\emph{\JournalTitle{Mechanics of Materials}}}
  \textbf{\bibinfo{volume}{182}}, \bibinfo{pages}{104668},
  \doiprefix\url{https://doi.org/10.1016/j.mechmat.2023.104668}
  (\bibinfo{year}{2023}).

\bibitem{reddy2006theory}
\bibinfo{author}{Reddy, J.~N.}
\newblock \emph{\bibinfo{title}{Theory and analysis of elastic plates and
  shells}} (\bibinfo{publisher}{CRC press}, \bibinfo{year}{2006}).

\bibitem{Vidyasagar2018}
\bibinfo{author}{Vidyasagar, A.}, \bibinfo{author}{Kr{\"o}del, S.} \&
  \bibinfo{author}{Kochmann, D.~M.}
\newblock \bibinfo{journal}{\bibinfo{title}{Microstructural patterns with
  tunable mechanical anisotropy obtained by simulating anisotropic spinodal
  decomposition}}.
\newblock {\emph{\JournalTitle{Proceedings of the Royal Society A:
  Mathematical, Physical and Engineering Sciences}}}
  \textbf{\bibinfo{volume}{474}}, \bibinfo{pages}{20180535}
  (\bibinfo{year}{2018}).

\bibitem{Lorensen1987}
\bibinfo{author}{Lorensen, W.~E.} \& \bibinfo{author}{Cline, H.~E.}
\newblock \bibinfo{journal}{\bibinfo{title}{Marching cubes: A high resolution
  3d surface construction algorithm}}.
\newblock {\emph{\JournalTitle{SIGGRAPH Comput. Graph.}}}
  \textbf{\bibinfo{volume}{21}}, \bibinfo{pages}{163–169},
  \doiprefix\url{10.1145/37402.37422} (\bibinfo{year}{1987}).

\bibitem{Goldman2005}
\bibinfo{author}{Goldman, R.}
\newblock \bibinfo{journal}{\bibinfo{title}{Curvature formulas for implicit
  curves and surfaces}}.
\newblock {\emph{\JournalTitle{Computer Aided Geometric Design}}}
  \textbf{\bibinfo{volume}{22}}, \bibinfo{pages}{632--658},
  \doiprefix\url{https://doi.org/10.1016/j.cagd.2005.06.005}
  (\bibinfo{year}{2005}).
\newblock \bibinfo{note}{Geometric Modelling and Differential Geometry}.

\bibitem{CALLENS2020}
\bibinfo{author}{Callens, S.~J.}, \bibinfo{author}{Uyttendaele, R.~J.},
  \bibinfo{author}{Fratila-Apachitei, L.~E.} \& \bibinfo{author}{Zadpoor,
  A.~A.}
\newblock \bibinfo{journal}{\bibinfo{title}{Substrate curvature as a cue to
  guide spatiotemporal cell and tissue organization}}.
\newblock {\emph{\JournalTitle{Biomaterials}}} \textbf{\bibinfo{volume}{232}},
  \bibinfo{pages}{119739},
  \doiprefix\url{https://doi.org/10.1016/j.biomaterials.2019.119739}
  (\bibinfo{year}{2020}).

\bibitem{Yang2022}
\bibinfo{author}{Yang, Y.} \emph{et~al.}
\newblock \bibinfo{journal}{\bibinfo{title}{Gaussian curvature–driven
  direction of cell fate toward osteogenesis with triply periodic minimal
  surface scaffolds}}.
\newblock {\emph{\JournalTitle{Proceedings of the National Academy of
  Sciences}}} \textbf{\bibinfo{volume}{119}}, \bibinfo{pages}{e2206684119},
  \doiprefix\url{10.1073/pnas.2206684119} (\bibinfo{year}{2022}).
\newblock \eprint{https://www.pnas.org/doi/pdf/10.1073/pnas.2206684119}.

\bibitem{tozzi2017}
\bibinfo{author}{Tozzi, G.} \emph{et~al.}
\newblock \bibinfo{journal}{\bibinfo{title}{Strain uncertainties from two
  digital volume correlation approaches in prophylactically augmented
  vertebrae: Local analysis on bone and cement-bone microstructures}}.
\newblock {\emph{\JournalTitle{Journal of the mechanical behavior of biomedical
  materials}}} \textbf{\bibinfo{volume}{67}}, \bibinfo{pages}{117--126}
  (\bibinfo{year}{2017}).

\bibitem{leoni1998}
\bibinfo{author}{Leoni, S.}
\newblock \emph{\bibinfo{title}{Applications of periodic surfaces for the study
  of crystal structures, first order phase transitions, force ordering, and
  systematic generation of novel structure models}}.
\newblock Ph.D. thesis, \bibinfo{school}{ETH Zurich} (\bibinfo{year}{1998}).

\bibitem{GANDY2001}
\bibinfo{author}{Gandy, P.~J.}, \bibinfo{author}{Bardhan, S.},
  \bibinfo{author}{Mackay, A.~L.} \& \bibinfo{author}{Klinowski, J.}
\newblock \bibinfo{journal}{\bibinfo{title}{Nodal surface approximations to the
  p,g,d and i-wp triply periodic minimal surfaces}}.
\newblock {\emph{\JournalTitle{Chemical Physics Letters}}}
  \textbf{\bibinfo{volume}{336}}, \bibinfo{pages}{187--195},
  \doiprefix\url{https://doi.org/10.1016/S0009-2614(00)01418-4}
  (\bibinfo{year}{2001}).

\bibitem{karcher1996}
\bibinfo{author}{Karcher, H.} \& \bibinfo{author}{Polthier, K.}
\newblock \bibinfo{journal}{\bibinfo{title}{Construction of triply periodic
  minimal surfaces}}.
\newblock {\emph{\JournalTitle{Philosophical Transactions of the Royal Society
  of London. Series A: Mathematical, Physical and Engineering Sciences}}}
  \textbf{\bibinfo{volume}{354}}, \bibinfo{pages}{2077--2104}
  (\bibinfo{year}{1996}).

\bibitem{meeks1990}
\bibinfo{author}{Meeks~III, W.~H.}
\newblock \bibinfo{journal}{\bibinfo{title}{The theory of triply periodic
  minimal surfaces}}.
\newblock {\emph{\JournalTitle{Indiana University Mathematics Journal}}}
  \bibinfo{pages}{877--936} (\bibinfo{year}{1990}).

\bibitem{YANG2010}
\bibinfo{author}{Yang, S.-D.}, \bibinfo{author}{Lee, H.~G.} \&
  \bibinfo{author}{Kim, J.}
\newblock \bibinfo{journal}{\bibinfo{title}{A phase-field approach for
  minimizing the area of triply periodic surfaces with volume constraint}}.
\newblock {\emph{\JournalTitle{Computer Physics Communications}}}
  \textbf{\bibinfo{volume}{181}}, \bibinfo{pages}{1037--1046},
  \doiprefix\url{https://doi.org/10.1016/j.cpc.2010.02.010}
  (\bibinfo{year}{2010}).

\bibitem{nesper2001}
\bibinfo{author}{Nesper, R.} \& \bibinfo{author}{Leoni, S.}
\newblock \bibinfo{journal}{\bibinfo{title}{On tilings and patterns on
  hyperbolic surfaces and their relation to structural chemistry}}.
\newblock {\emph{\JournalTitle{ChemPhysChem}}} \textbf{\bibinfo{volume}{2}},
  \bibinfo{pages}{413--422} (\bibinfo{year}{2001}).

\bibitem{baena2021}
\bibinfo{author}{Baena-Moreno, F.~M.} \emph{et~al.}
\newblock \bibinfo{journal}{\bibinfo{title}{Stepping toward efficient
  microreactors for co2 methanation: 3d-printed gyroid geometry}}.
\newblock {\emph{\JournalTitle{ACS Sustainable Chemistry \& Engineering}}}
  \textbf{\bibinfo{volume}{9}}, \bibinfo{pages}{8198--8206}
  (\bibinfo{year}{2021}).

\bibitem{kapfer2011}
\bibinfo{author}{Kapfer, S.~C.}, \bibinfo{author}{Hyde, S.~T.},
  \bibinfo{author}{Mecke, K.}, \bibinfo{author}{Arns, C.~H.} \&
  \bibinfo{author}{Schr{\"o}der-Turk, G.~E.}
\newblock \bibinfo{journal}{\bibinfo{title}{Minimal surface scaffold designs
  for tissue engineering}}.
\newblock {\emph{\JournalTitle{Biomaterials}}} \textbf{\bibinfo{volume}{32}},
  \bibinfo{pages}{6875--6882} (\bibinfo{year}{2011}).

\bibitem{bonatti2019}
\bibinfo{author}{Bonatti, C.} \& \bibinfo{author}{Mohr, D.}
\newblock \bibinfo{journal}{\bibinfo{title}{Mechanical performance of
  additively-manufactured anisotropic and isotropic smooth shell-lattice
  materials: Simulations \& experiments}}.
\newblock {\emph{\JournalTitle{Journal of the Mechanics and Physics of
  Solids}}} \textbf{\bibinfo{volume}{122}}, \bibinfo{pages}{1--26}
  (\bibinfo{year}{2019}).

\bibitem{ross1992}
\bibinfo{author}{Ross, M.}
\newblock \bibinfo{journal}{\bibinfo{title}{Schwarz'p and d surfaces are
  stable}}.
\newblock {\emph{\JournalTitle{Differential Geometry and its Applications}}}
  \textbf{\bibinfo{volume}{2}}, \bibinfo{pages}{179--195}
  (\bibinfo{year}{1992}).

\bibitem{guo2022}
\bibinfo{author}{Guo, X.} \emph{et~al.}
\newblock \bibinfo{journal}{\bibinfo{title}{Enhancement in the mechanical
  behaviour of a schwarz primitive periodic minimal surface lattice structure
  design}}.
\newblock {\emph{\JournalTitle{International Journal of Mechanical Sciences}}}
  \textbf{\bibinfo{volume}{216}}, \bibinfo{pages}{106977}
  (\bibinfo{year}{2022}).

\bibitem{ciliberto2015}
\bibinfo{author}{Ciliberto, C.}, \bibinfo{author}{Lopes, M.~M.} \&
  \bibinfo{author}{Roulleau, X.}
\newblock \bibinfo{journal}{\bibinfo{title}{On schoen surfaces}}.
\newblock {\emph{\JournalTitle{Commentarii Mathematici Helvetici}}}
  \textbf{\bibinfo{volume}{90}}, \bibinfo{pages}{59--74}
  (\bibinfo{year}{2015}).

\bibitem{rito2016}
\bibinfo{author}{Rito, C.}, \bibinfo{author}{Roulleau, X.} \&
  \bibinfo{author}{Sarti, A.}
\newblock \bibinfo{journal}{\bibinfo{title}{Explicit schoen surfaces}}.
\newblock {\emph{\JournalTitle{arXiv preprint arXiv:1609.02235}}}
  (\bibinfo{year}{2016}).

\bibitem{bathe1986formulation}
\bibinfo{author}{Bathe, K.-J.} \& \bibinfo{author}{Dvorkin, E.~N.}
\newblock \bibinfo{journal}{\bibinfo{title}{A formulation of general shell
  elements—the use of mixed interpolation of tensorial components}}.
\newblock {\emph{\JournalTitle{International journal for numerical methods in
  engineering}}} \textbf{\bibinfo{volume}{22}}, \bibinfo{pages}{697--722}
  (\bibinfo{year}{1986}).

\bibitem{Shevchenko2023}
\bibinfo{author}{Shevchenko, V.}, \bibinfo{author}{Balabanov, S.},
  \bibinfo{author}{Sychov, M.} \& \bibinfo{author}{Karimova, L.}
\newblock \bibinfo{journal}{\bibinfo{title}{Prediction of cellular structure
  mechanical properties with the geometry of triply periodic minimal surfaces
  (tpms)}}.
\newblock {\emph{\JournalTitle{ACS Omega}}} \textbf{\bibinfo{volume}{8}},
  \bibinfo{pages}{26895--26905}, \doiprefix\url{10.1021/acsomega.3c01631}
  (\bibinfo{year}{2023}).
\newblock \eprint{https://doi.org/10.1021/acsomega.3c01631}.

\bibitem{MASKERY2018}
\bibinfo{author}{Maskery, I.} \emph{et~al.}
\newblock \bibinfo{journal}{\bibinfo{title}{Insights into the mechanical
  properties of several triply periodic minimal surface lattice structures made
  by polymer additive manufacturing}}.
\newblock {\emph{\JournalTitle{Polymer}}} \textbf{\bibinfo{volume}{152}},
  \bibinfo{pages}{62--71},
  \doiprefix\url{https://doi.org/10.1016/j.polymer.2017.11.049}
  (\bibinfo{year}{2018}).
\newblock \bibinfo{note}{SI: Advanced Polymers for 3DPrinting/Additive
  Manufacturing}.

\bibitem{QIU2023}
\bibinfo{author}{Qiu, N.}, \bibinfo{author}{Wan, Y.}, \bibinfo{author}{Shen,
  Y.} \& \bibinfo{author}{Fang, J.}
\newblock \bibinfo{journal}{\bibinfo{title}{Experimental and numerical studies
  on mechanical properties of tpms structures}}.
\newblock {\emph{\JournalTitle{International Journal of Mechanical Sciences}}}
  \bibinfo{pages}{108657},
  \doiprefix\url{https://doi.org/10.1016/j.ijmecsci.2023.108657}
  (\bibinfo{year}{2023}).

\bibitem{SchrderTurk2011}
\bibinfo{author}{Schr\"{o}der-Turk, G.~E.} \emph{et~al.}
\newblock \bibinfo{journal}{\bibinfo{title}{Minkowski tensor shape analysis of
  cellular, granular and porous structures}}.
\newblock {\emph{\JournalTitle{Advanced Materials}}}
  \textbf{\bibinfo{volume}{23}}, \bibinfo{pages}{2535--2553},
  \doiprefix\url{10.1002/adma.201100562} (\bibinfo{year}{2011}).

\bibitem{paszke2019pytorch}
\bibinfo{author}{Paszke, A.} \emph{et~al.}
\newblock \bibinfo{journal}{\bibinfo{title}{Pytorch: An imperative style,
  high-performance deep learning library}}.
\newblock {\emph{\JournalTitle{Advances in neural information processing
  systems}}} \textbf{\bibinfo{volume}{32}} (\bibinfo{year}{2019}).

\bibitem{kingma2014adam}
\bibinfo{author}{Kingma, D.~P.} \& \bibinfo{author}{Ba, J.}
\newblock \bibinfo{journal}{\bibinfo{title}{Adam: A method for stochastic
  optimization}}.
\newblock {\emph{\JournalTitle{arXiv preprint arXiv:1412.6980}}}
  (\bibinfo{year}{2014}).

\end{thebibliography}

\newpage

\setcounter{section}{0}
\setcounter{figure}{0}
\setcounter{table}{0}
\renewcommand{\figurename}{Supporting Figure}
\renewcommand{\tablename}{Supporting Table}

{\LARGE \noindent\textbf{Supporting Information}}

\vskip 15pt

{\Large \noindent\textbf{Inverse designing surface curvatures by deep learning}}

\vskip 15pt

{\large \noindent\textbf{Yaqi Guo$^{1,2}$,\  Saurav Sharma$^{1*}$,\  Siddhant Kumar$^{1*}$}}

\vskip 10pt

\noindent $^1$ Department of Materials Science and Engineering, Delft University of Technology, Netherlands

\noindent $^2$ School of Aerospace Engineering and Applied Mechanics, Tongji University, China

\noindent $^*$ S.Sharma-7@tudelft.nl,\ Sid.Kumar@tudelft.nl

\vskip 30pt

\section{Geometric interpretation of the design space}
The standard design parameters $\bfTheta$ in the curvature-based energy functional can be obtained in terms of the geometrically interpretable but equivalent parameters $\tilde \bfTheta$ as:
\be
\begin{aligned}
	a_{20} &= \frac{g(1+\alpha-\alpha\cos{2\theta}+\cos{2\theta})}{2}\\
	a_{11} &= g(1-\alpha)\sin{2\theta}\\
	a_{02} &= \frac{g(1+\alpha+\alpha\cos{2\theta}-\cos{2\theta})}{2}\\
	a_{10} &= -g\bigl[(1+\alpha)\kappa^c_1+(1-\alpha)\kappa^c_1\cos{2\theta}+(1-\alpha)\kappa^c_2\sin{2\theta}\bigr]\\
	a_{01} &= -g\bigl[(1+\alpha)\kappa^c_2+(\alpha-1)\kappa^c_2\cos{2\theta}+(1-\alpha)\kappa^c_1\sin{2\theta}\bigr]\\
	a_{00} &= \frac{g\bigl[(1+\alpha)(\kappa^{c2}_1+\kappa^{c2}_2)+(1-\alpha)(\kappa^{c2}_1-\kappa^{c2}_2)\cos{2\theta}+2\kappa^{c}_1\kappa^{c}_2(1-\alpha)\sin{2\theta}-2c\bigr]}{2}
\end{aligned}
\ee

\section{Data sampling and scaling strategy}

\begin{figure}
	\includegraphics[width=\linewidth]{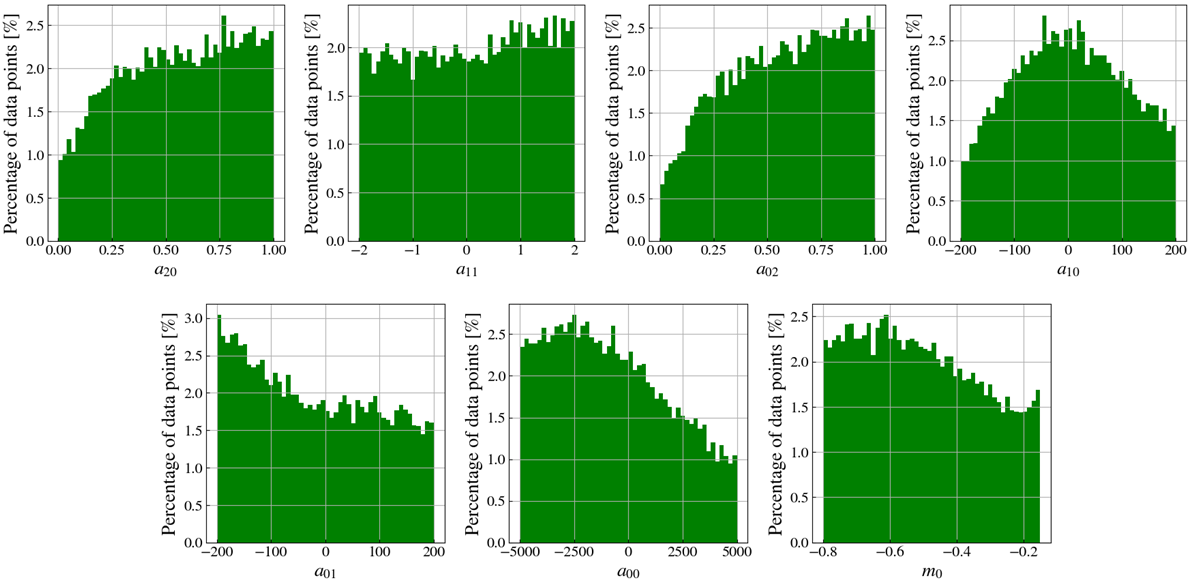}
	\caption{Distribution of components of $\bfTheta$ in the dataset before scaling.}
	\label{fig:dist-theta}
\end{figure}

To generate the training dataset, we randomly sample the design parameters $\bfTheta$. However, some combinations of the design parameters may lead to energetics that do not result in phase separation, or the convergence of the phase-field approximation might not be achieved, rendering not all possible $\bfTheta$ physically valid\cite{Song2021}. Through trial-and-error with $1000$ attempts, we identify that the domain ${a_{20}\in[0,1]}$, ${a_{11}\in[-2,2]}$, ${a_{02}\in[0,1]}$, ${a_{10}\in[-200,200]}$, ${a_{01}\in[-200,200]}$, ${a_{00}\in[-5000,5000]}$ and ${m_0}\in[-0.8,-0.15]$ provides feasible energetics with high probability and sufficient diversity. We perform  uniform random sampling of $\bfTheta$ in this domain and discarding those which result in infeasible energetics, following which the topology and curvature profile (equivalently, curvature encodings $\bfchi$) are computed using the open-source implementation of ref. \cite{Song2021} \figurename~\ref{fig:dist-theta} illustrates the distribution of the components of $\bfTheta$ in the dataset.

\begin{figure}
	\includegraphics[width=\textwidth]{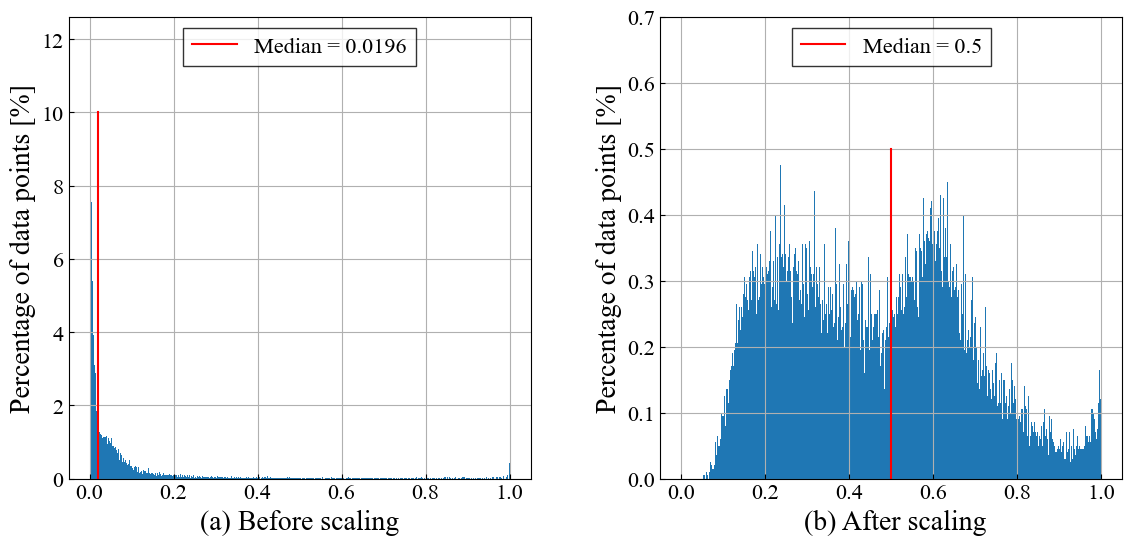}
	\caption{Distribution of maximal value of each component of curvature encodings in the dataset: (a) before and (b) after scaling.}
	\label{fig:dist-chi}
\end{figure}

For efficient ML model training, we independently scale each design parameter in $\bfTheta\in\Rset^7$ using their corresponding minimum and maximum values in the training dataset as
\be\label{eq:theta-scaling}
\Theta_j^\text{s} = \frac{\Theta_j-\min_{i\in \calD}\Theta^{(i)}_j}{\max_{i\in \calD}\Theta^{(i)}_j-\min_{i\in \calD}\Theta^{(i)}_j},\qquad \forall j=1\dots 7.
\ee
The curvature encodings in $\bfchi\in[0,1]^k$, however, can be highly sparse and exhibit a distribution that is heavily skewed near the values equal to zero and one(see \figurename~\ref{fig:dist-chi}a). Therefore, we perform an independent nonlinear scaling for each component given by
\be\label{eq:chi-scaling}
\chi_j^\text{s} = h(\chi_j) \qquad \text{with}\qquad  h(\chi_j)= \frac{\ln{(\chi_j-a)}}{b}+c,\qquad \forall j=1,\dots,k,
\ee
where the constants ${a,b,c}$ are uniquely determined such that 
\be
h(0) = 0, 
\qquad
h\left(\text{Median}\left(\left\{\max_{i\in \calD}\chi^{(i)}_j: j=1,\dots,k\right\}\right)\right)= 0.5, 
\qquad 
h(1) = 1.
\ee
The scaling ensures that the lower and upper bound of the data remain at zero or one, while the median of the maximum value of each curvature encoding over the whole dataset is mapped to a value of $0.5$. After scaling, the small values near zero get raised and the distribution becomes significantly less skewed (see \figurename~\ref{fig:dist-chi}b). The scaled design parameters and curvature encodings, i.e., $\bfTheta^\text{s}$ and $\bfchi^\text{s}$ are then used for training of the ML models.

\section{Machine learning protocols and computing times}

\tablename~\ref{tab:protocols} summarizes the protocols (such as learning rate, optimizer, batch size etc.) for training of the \textit{f}-NN and \textit{i}-NN. The \textit{f}-NN and \textit{i}-NN consist of several hidden layers of different dimensions with each followed by an activation function (see \tablename~\ref{tab:protocols} and \figurename~\ref{fig:nn-architecture}). Here, we use the rectified linear unit \cite{paszke2019pytorch} (ReLU) as the activation function ($\text{ReLU}(\cdot)=\max(0,\cdot)$). Since each component of both the normalized design parameters $\bfTheta^\text{s}$ and normalized curvature encoding $\bfchi^\text{s}$ is bounded in $[0,1]$, we use a scaled version of ReLU6 \cite{paszke2019pytorch}, i.e., $\text{ReLU6}(\cdot)/6=\min(\max(0,\cdot),6)/6$ as the last activation layer (before output) for both \textit{i}-NN and \textit{f}-NN. This serves as an inductive bias by identically satisfying the output bounds of both the NNs and therefore, improves the training efficiency.

\tablename~\ref{tab:compute} presents the computing times as well as  software and hardware used for different tasks. Notably, the fast prediction time of the surrogate \textit{f}-NN compared to the phase-field model is key to making the \textit{i}-NN training possible. Despite the one-time cost of data generation and training, both \textit{f}-NN and \textit{i}-NN  -- once trained -- provide instant exploration of both the design-to-curvature and curvature-to-design maps.

{For all the pairs of design parameters and curvature encodings in the test set, \figurename~\ref{fig:R2-SI} compares the true design parameters and the predicted design parameters (obtained from the \textit{i}-NN with the corresponding curvature encoding as input). We \textit{expectedly} observe a poor correlation between the ground truth and predictions, which highlights the ill-posedness of the inverse problem  (i.e., multiple designs parameters can lead to similar curvature profiles) and the advantage of the \textit{i}-NN training strategy.}

\begin{figure}
	\includegraphics[width=\textwidth]{nn-architecture.png}
	\caption{Architectures of the \textbf{(a)} ${f}$-NN and \textbf{(b)} ${i}$-NN models. The dimension of each input, output, or hidden layer is denoted by the number above it.}
	\label{fig:nn-architecture}
\end{figure}

\begin{figure}
	\includegraphics[width=\textwidth]{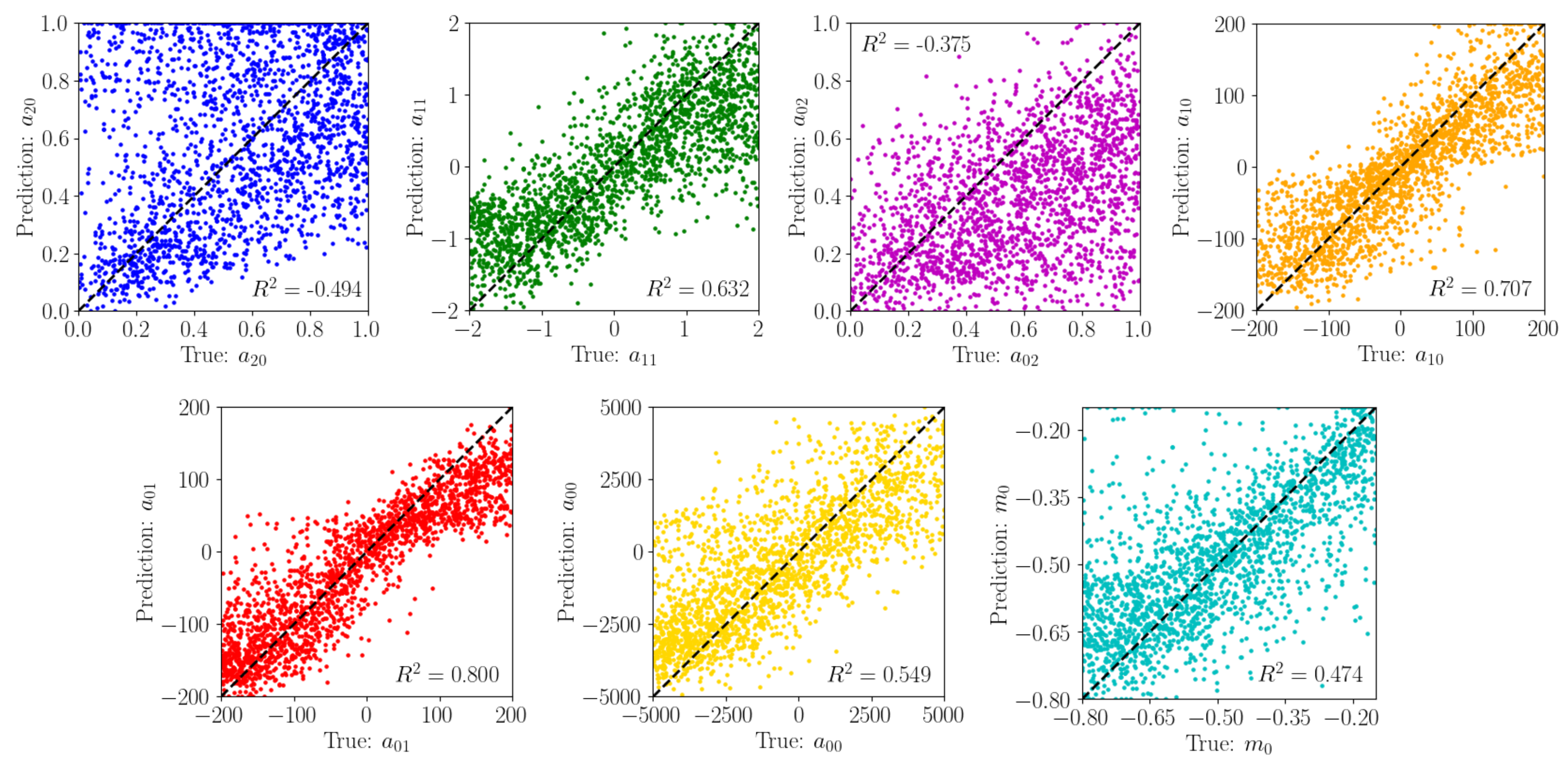}
	\caption{{Predicted (via the \textit{i}-NN) vs.~true design parameters $\bfTheta$ in the test dataset. All dashed lines represent the ideal line with zero-intercept and unit-slope; the corresponding $R^2$ goodness-of-fit are indicated. As expected, the poor correlations between the true and predicted design parameters highlight the ill-posed nature of the inverse design problem and the robustness of the \textit{i}-NN training strategy that enables achieving accurate curvature profiles despite the mismatch in the underlying design parameters.}}
	\label{fig:R2-SI}
\end{figure}

\begin{table}
	\centering
	\caption{Training protocols for the ${f}$-NN and ${i}$-NN models.}
	\begin{tabular}{ c c c }
		\hline
		& ${f}$-NN & ${i}$-NN \\ 
		\hline
		Hidden layer dimensions & 50, 150, 300, 600, 1200, 2500, 5000, 10000, 20000 & 5000, 1000, 200, 40 \\ 
		%   \hline
		Activation functions & ReLU6/6 for output layer; ReLU otherwise & ReLU6/6 for output layer; ReLU otherwise \\
		Input data scaling & min-max (see \eqref{eq:theta-scaling}) & custom nonlinear scaling (see \eqref{eq:chi-scaling}) \\ 
		Optimizer & Adam \cite{kingma2014adam} & Adam \cite{kingma2014adam} \\ 
		Learning rate & ${10^{-4}}$ & ${10^{-4}}$ \\ 
		Batch size & 128 & 128 \\
		Number of epochs & 220 & 600 \\
		Dropout & none & none \\
		\hline
	\end{tabular}
	% \caption{Training protocols for the ${f}$-NN and ${i}$-NN models.}
	\label{tab:protocols}
\end{table}

\begin{table}
	\begin{center}
		\caption{Task-wise breakdown of compute runtime, software, and hardware used.}
		\begin{tabular}{ c c c c }
			\hline
			Task & Software & Hardware & Runtime \\ 
			\hline
			Generating one topology using the phase-field model & PyTorch &  CPU+GPU  & {2 minutes per sample}\\ 
			Training ${f}$-NN & PyTorch  & CPU+GPU  & {1.5 hours} \\ 
			Training ${i}$-NN & PyTorch  & CPU+GPU & {3 hours} \\ 
			Design parameters prediction using ${i}$-NN & PyTorch  & CPU+GPU  & {<1 second} \\ 
			\hline
		\end{tabular}
		% \caption{Task-wise breakdown of compute runtime, software, and hardware used. We note that the runtimes reported are rough estimates only and may vary across different computing hardware. Here, we used two 3.0 GHz 24-core Intel Xeon Gold 6248R processors with combined 256GB DDR4 memory and a Nvidia RTX A5000 GPU with 24GB memory.}
		\label{tab:compute}
	\end{center}
	\vspace{1ex}
	{ We note that the runtimes reported are rough estimates only and may vary across different computing hardware. Here, we used two 3.0 GHz 24-core Intel Xeon Gold 6248R processors with combined 256GB DDR4 memory and a Nvidia RTX A5000 GPU with 24GB memory. \par}
\end{table}

\section{Processing 3D-voxelated image data of trabecular bone}

Benchmark 1 (see main article) tests the generalization capability of the inverse design framework on trabecular bone. However, the raw data of trabecular bone from micro-computed tomography is in the form of voxelization-based 3D image, which is too rough for estimating the surface curvature profile. To obtain a smooth surface, we first apply a min-max scaling on the grayscale 3D image, followed by a 3D transposed convolution and a 3D convolution operation. The latter are implemented via PyTorch functions with the following parameters:
\begin{itemize}
	\item Transposed convolution: \texttt{ConvTranspose3d()} with \\ kernel \texttt{size=(4,4,4)}, \texttt{stride=(2,2,2)}, and \texttt{padding=(1,1,1)};
	\item Convolution: \texttt{Conv3d()} with \\ kernel \texttt{size=(3,3,3)}, \texttt{stride=(2,2,2)}, and \texttt{padding=(1,1,1)}.
\end{itemize}
All the kernel weights (in both convolution layers) are set equal to one, which -- in addition to small kernel size, stride, and padding -- effectively applies localized smoothing of the solid-void interface. The resulting smoothed data is then passed through a hyperbolic tangent transformation, followed by the application of a zero level set to compute the trabecular bone surfaces.

\section{Additional benchmark results}

\begin{figure}[t]
	\includegraphics[width=\linewidth]{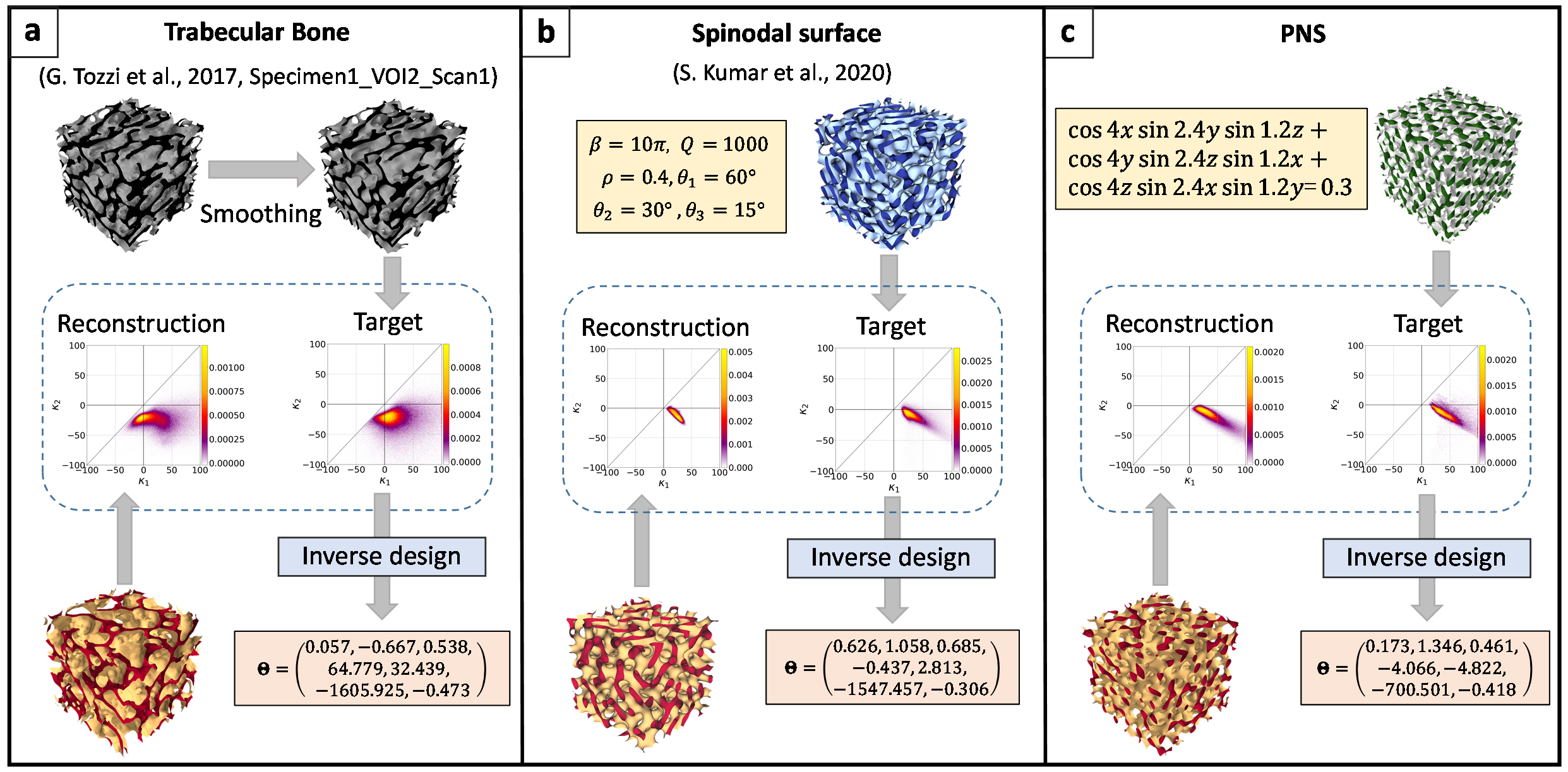}
	\caption{Additional inverse design benchmarks including (a) trabecular bone, (b) spinodoid structure, and (c) PNS surface.}
	\label{fig:additional-application}
\end{figure}

In addition to Figure 7 of the main article, here we present another set of inverse design benchmarks in \figurename~\ref{fig:additional-application}  -- including trabecular bone, spinodoid structure, and PNS surface. All the parameters and data processing protocols remain the same as those discussed in the main article, with the exception of only the parameters explicitly mentioned in \figurename~\ref{fig:additional-application}. The observations regarding the performance and generalization of the \textit{i}-NN here are consistent with the observations in the main article.

\section{{Strain energy distribution under shear load}}
\begin{figure}
	\centering
	\includegraphics[width=\textwidth, height = \dimexpr \textheight - 4\baselineskip\relax, keepaspectratio]{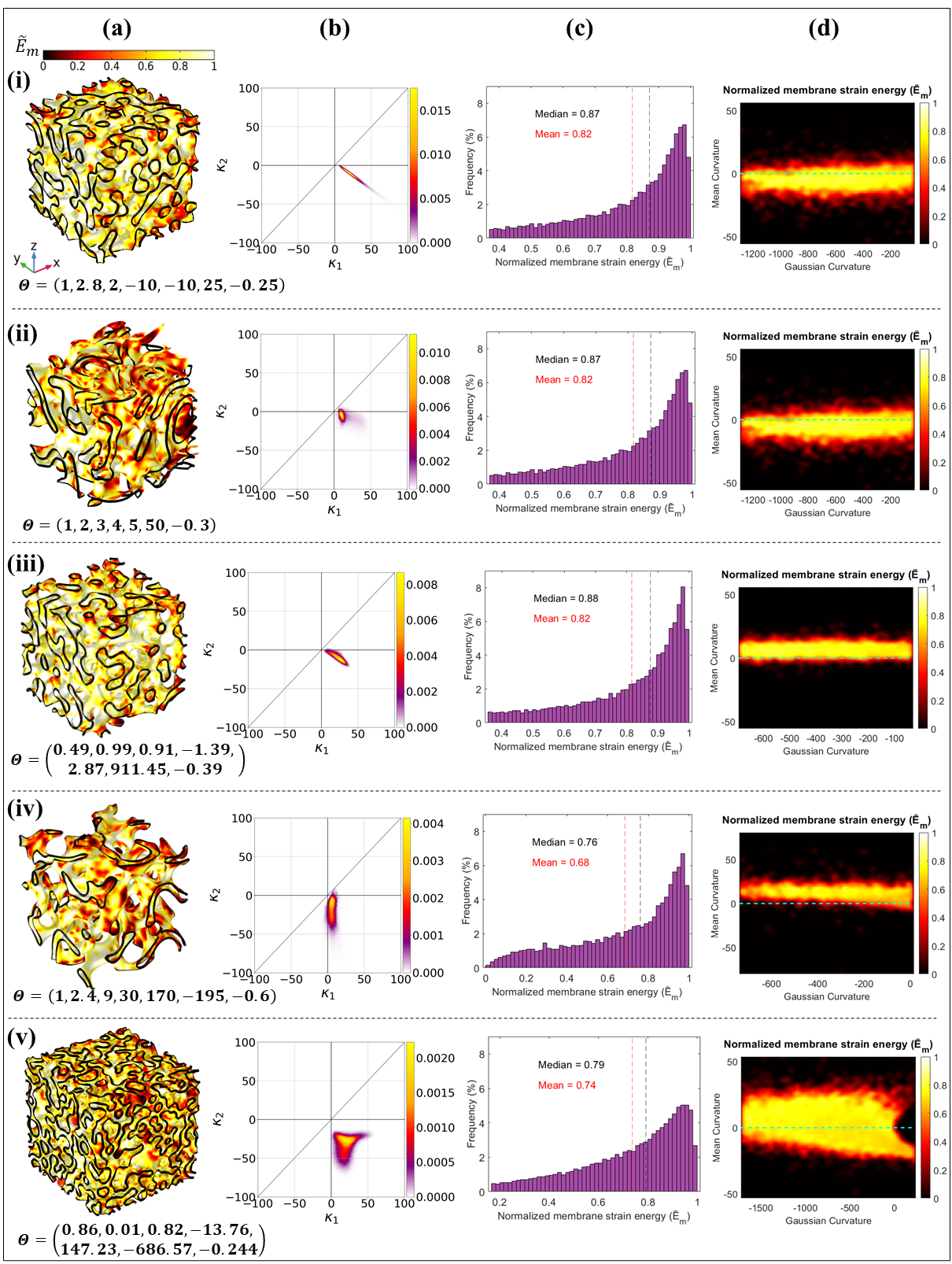}
	\caption{{Interplay of bending and membrane strain energy in the five diverse topologies (topologies (i) to (v) of the Figure 8) under unit shear load. a) Distribution of normalized membrane strain energy $\tilde{E_m}$ in the topologies and (b) their corresponding curvature profiles. (c) Distribution of $\tilde{E_m}$ over the surface area of each topology. (d) Distribution of mean and Gaussian curvatures; colored by the mean of $\tilde{E_m}$ for the corresponding curvature values found within a topology.}}
	\label{fig:mechanics_shear}
\end{figure}

{The distribution of strain energy in the form of membrane and bending energy under shear load, i.e., unit load applied to the face $X=100$ in $z$-direction and fixed constraint on $X=0$, is plotted in Figure \figurename~\ref{fig:mechanics_shear}. The spatial distribution of the normalized membrane strain energy shows a trend similar to that of Figure 8. Therefore, the correlations between curvature profile and stretching vs.~bending dominated behavior remain the same across different loading conditions.}

\end{document}